\documentclass[11pt]{article}
\usepackage{amsmath}
\usepackage{amsfonts}
\usepackage{ipa}
\def\p{\partial}
\def\be{\begin{equation}}
\def\ee{\end{equation}}
\def\bea{\begin{eqnarray}}
\def\eea{\end{eqnarray}}
\def\bea*{\begin{eqnarray*}}
\def\eea*{\end{eqnarray*}}
\def\scri{\mathcal{I}^+}

\def\hF{\hat{F}}

\def\hm{\hat{m}}
\def\hg{\hat{g}}

\def\hV{\hat{V}}

\def\hT{\hat{T}}

\def\hGamma{\hat{\Gamma}}

\def\cF{\check{F}}

\def\cg{\check{g}}
\def\cm{\check{m}}
\def\cM{\check{M}}

\def\cT{\check{T}}

\def\cC{\check{C}}

\def\cOmega{\check{\Omega}}

\def\crho{\check{\rho}}

\def\cu{\check{u}}
\def\ca{\check{a}}
\def\cm{\check{m}}
\def\ba{\bf{a}}
\def\bb{\bf{b}}
\def\bc{\bf{c}}
\def\bd{\bf{d}}
\def\scri{{\mathcal{I}}}

\begin{document}
\title{Conformal methods in General Relativity with application to Conformal Cyclic Cosmology: \newline A minicourse at IX IMLG}
\author{Paul Tod}
\date{June 18th-22nd 2018}
\maketitle
\begin{abstract}
In these lectures my aim is to review enough of conformal differential geometry in four dimensions to give an account of Penrose's conformal cyclic cosmology.
\end{abstract}
\section{Tensor calculus and conformal rescaling}
\label{S1}
My approach will be fairly concrete. For more abstract approaches see \cite{cg1} or \cite{eas}.
\subsection{Differential geometry conventions}
\label{SS1.1}
These largely follow \cite{pr}: we have a metric $g_{ab}$ with signature $(+---)$ and will use abstract indices where possible. The Christoffel symbols, which need concrete indices, here written as bold, are
\be\label{1} \Gamma^{\ba}_{\bb\bc}=\frac12g^{\ba\bd}(g_{\bd\bb,\bc}+g_{\bd\bc,\bb}-g_{\bb\bc,\bd}),\ee
and the metric or Levi-Civita covariant derivative in concrete indices is
\[\nabla_{\ba} X^{\bb}=\frac{\p}{\p x^{\ba}}X^{\bb}+\Gamma^{\bb}_{\ba\bc}X^{\bc}.\]
The Riemann tensor is defined in abstract indices by
\be\label{2}
(\nabla_a\nabla_b-\nabla_b\nabla_a)X^c=R_{abd}^{\;\;\;\;\;c}X^d.\ee
It decomposes into irreducible parts as
\be\label{3}
R_{abcd}=C_{abcd}+\frac12(g_{ac}R_{bd}+g_{bd}R_{ac}-g_{bc}R_{ad}-g_{ad}R_{bc})-\frac16R(g_{ac}g_{bd}-g_{ad}g_{bc}),\ee
with $C_{abcd}$ the Weyl tensor and
\[R_{ac}=R_{adc}^{\;\;\;\;\;d},\;\;R=g^{ab}R_{ab},\]
the Ricci tensor and Ricci scalar respectively. The Einstein tensor is 
\[G_{ab}=R_{ab}-\frac12Rg_{ab},\]
and with the conventions used here, the Einstein field equations (or EFEs) with cosmological constant $\lambda$ are
\be\label{efe}
G_{ab}=-8\pi GT_{ab}-\lambda g_{ab},\ee
and for vacuum
\be\label{efe2}R_{ab}=\lambda g_{ab}.\ee

\medskip

We recall the definition
\be\label{8}
L_{ab}=-\frac12(R_{ab}-\frac16Rg_{ab})\ee
of a tensor sometimes called the \emph{Schouten tensor}, which will be useful below. It is called $P_{ab}$ by some authors (where the kernel letter is a capital rho), and allows a slight simplification of (\ref{3}):
\[R_{ab}^{\;\;\;\;cd}=C_{ab}^{\;\;\;\;cd}-4\delta_{[a}^{\;\;[c}L_{b]}^{\;\;d]}.\]

For the record, recall the Bianchi identity in the form
\be\label{9} \nabla_dC_{abc}^{\;\;\;\;\;d}=2\nabla_{[a}L_{b]c}.\ee
The tensor on the right here is sometimes called the \emph{Cotton tensor}.

\medskip

The Lie derivative of a tensor by a given vector field $X^a$ is obtained recursively by 
\[\mathcal{L}_Xf=X^a\nabla_af,\;\;{\mathcal{L}}_XY^a=X^b\nabla_bY^a-Y^b\nabla_bX^a=[X,Y]^a\]
for a function $f$ and a vector field $Y$, where $[X,Y]^a$ is the Lie bracket, and then extended so as to satisfy Leibniz rule. It is independent of torsion-free connection, and in particular therefore
\[\mathcal{L}_Xg_{ab}= X^c\nabla_cg_{ab}+g_{ac}\nabla_bX^c+g_{bc}\nabla_aX^c=\nabla_aX_b+\nabla_bX_a.\]

A vector field $X^a$ is \emph{Killing} if $\mathcal{L}_Xg_{ab}= 0$ and \emph{conformal Killing} if $\mathcal{L}_Xg_{ab}= \phi g_{ab}$ for some function $\phi$. The importance of such vector fields is in defining 
conserved quantities for the geodesic equation: 
\begin{itemize}
 \item 
Recall 
a curve $\Gamma$ with tangent vector $V^a$ is a \emph{geodesic} if 
\[V^b\nabla_bV^a=fV^a,\]
and affinely-parametrised if this holds with $f=0$. 
\item A geodesic is time-like, space-like or null according as $g_{ab}V^aV^b$ is positive, negative or zero (and of course this holds at all points of $\Gamma$ if it holds 
at one). 
\item For an affinely-parametrised geodesic $\Gamma$ with tangent vector $V^a$ and a Killing vector $X^a$, the contraction $g_{ab}V^aX^b$ is constant along $\Gamma$.  If 
the geodesic is null this is also true with a conformal Killing vector. 
\item One can define Killing tensors and conformal Killing tensors as tensors $K_{a_1\ldots a_n}$ which give rise to higher-order constants of 
the motion $K_{a_1\ldots a_n}V^{a_1}\ldots V^{a_n}$ for geodesics or null geodesics.
\item This discussion can be phrased in terms of the phase space or cotangent bundle $T^*(M)=\{(p_{a},q^{b})\}$ of space-time, where the indices can be taken to be abstract, 
and the Poisson bracket $\{\;\;,\;\;\}$ on $T^*(M)$: there is a vector field 
$\{g^{ab}p_ap_b,\cdot\}$ 
on $T^*(M)$ called the \emph{geodesic spray}, whose integral curves are lifted geodesics. A Killing vector defines a linear polynomial in momentum $X^ap_a$ which is constant along the geodesic spray:
\[\{g^{ab}p_ap_b,X^cp_c\}=0.\]
This language is useful in discussing the Einstein-Vlasov system.

\end{itemize}
\subsection{Conformal rescaling}
\label{SS1.2}
We shall be interested in conformally rescaling a space-time metric, which is the transformation
\be\label{2.1}g_{ab}\rightarrow\hg_{ab}=\Omega^2g_{ab},\ee
for a real-valued, usually smooth function $\Omega$. Necessarily then
\[\hg^{ab}=\Omega^{-2}g^{ab}\] and it's easy to see that
\be\label{2.2}\hGamma^{a}_{bc}-\Gamma^{a}_{bc}=\delta^{a}_{\;{b}}\Upsilon_{c}+\delta^{a}_{\;{c}}\Upsilon_{b}-g_{bc}g^{ad}\Upsilon_{d},\ee
where $\Upsilon_a=\Omega^{-1}\nabla_a\Omega$ (and we're allowed abstract indices as the difference between two connections \emph{is} a tensor). 

\medskip

It is now straightforward if tedious to calculate the transformation of the curvature. One finds

\begin{eqnarray}
 \hat{C}_{abc}^{\;\;\;\;\;d}&=&C_{abc}^{\;\;\;\;\;d},\\\label{c1}
 \hat{R}_{ab}&=&R_{ab}+2\nabla_a\Upsilon_b-2\Upsilon_a\Upsilon_b+g_{ab}(\nabla_c\Upsilon^c+2\Upsilon_c\Upsilon^c),\\\label{c2}
 \hat{R}&=&\Omega^{-2}(R+6\frac{\Box\Omega}{\Omega}).\label{c3}
\end{eqnarray}
The Schouten tensor (\ref{8}) transforms as

\be\label{l4}\hat{L}_{ab}=L_{ab}-\nabla_a\Upsilon_b+\Upsilon_a\Upsilon_b-\frac12g_{ab}\Upsilon_c\Upsilon^c.\ee

It's worth noting the transformation of the volume form:
\[\epsilon_{abcd}\rightarrow\hat{\epsilon}_{abcd}=\Omega^4\epsilon_{abcd},\]
whence also
\be\label{d1}\hat{\epsilon}_{ab}^{\;\;\;\;cd}=\epsilon_{ab}^{\;\;\;\;cd}\ee
so that this tensor, which defines duality on 2-forms, is therefore conformally invariant. 

The tensor $\epsilon^{abcd}\epsilon_{pqrs}$ which defines the volume form on the phase space $T^*(M)$ is also conformally invariant.
\subsubsection{Conformal weight}\label{SSS1.2.1}
A scalar $\omega$ is said to have \emph{conformal weight} $k$ if it transforms as
\[\omega\rightarrow\hat{\omega}=\Omega^k\omega\mbox{  when } g_{ab}\rightarrow\hg_{ab}=\Omega^2g_{ab}.\]
Evidently scalars of conformal weight $k$ can be regarded as sections of a bundle $\mathcal{E}^k$ and one can define a weighted covariant derivative
\[\nabla^{(cw)}_a\omega:=(\nabla_a-k\Upsilon_a)\omega,\]
though we won't be using this much.

Conformally-weighted vectors and tensors are defined by taking tensor products with conformally-weighted scalars. Thus the metic $g_{ab}$ has conformal 
weight 2, the duality operator $\epsilon_{ab}^{\;\;\;\;cd}$ has conformal weight 0, and so on.
\subsubsection{Geodesics}\label{SSS1.2.2}
A few points about the transformation of geodesics under conformal rescaling:
\begin{itemize}
 \item A null geodesic for $g_{ab}$ continues to be a null geodesic for $\hg_{ab}$; affine-parametrisation is preserved if one sets $\hV^a=\Omega^{-2}V^a$ or equivalently $\hV_a=V_a$.
 \item Time-like or space-like geodesics are not preserved.
 \item A Killing vector $X^a$ for $g_{ab}$ is a Killing vector for $\hg_{ab}$ iff $X^a\Upsilon_a=0$, otherwise it is a conformal Killing vector. Conformal Killing vectors transform to conformal Killing vectors but Killing 
 tensors in general don't transform nicely.
 \end{itemize}
 \subsubsection{Important examples of rescaling}
 \label{SSS1.2.3}
 \begin{itemize}
 \item {\bf{de Sitter space}}: This is the space of constant curvature obtained as the hyperboloid
 \[T^2-X^2-Y^2-Z^2-W^2=-H^{-2}=\mbox{  constant}\]
 in the 5-dimensional Minkowski space with metric
 \[g=dT^2-dX^2-dY^2-dZ^2-dW^2.\]
 It's an exercise to introduce coordinates (start with $T=H^{-1}\sinh Ht$) and find the metric as
 \[g=dt^2-H^{-2}\cosh^2(Ht)(dr^2+\sin^2r(d\theta^2+\sin^2\theta d\phi^2)),\]
 and then it isn't difficult to discover that
 \[C_{abc}^{\;\;\;\;\;d}=0,\;\;\;R_{ab}=3H^2g_{ab}.\]
 Evidently the underlying manifold is $\mathbb{R}\times\mathbb{S}^3$ and the $\mathbb{R}$-factor can be compactified by conformal rescaling with $\Omega=H\mbox{sech}(Ht)$:
 \be\label{esc}\hg=\Omega^2g=d\tau^2-(dr^2+\sin^2r(d\theta^2+\sin^2\theta d\phi^2)),\ee
 where
 \[d\tau=\frac{Hdt}{\cosh{Ht}},\;\;\mbox{  and w.l.o.g. }e^{Ht}=\tan(\tau/2),\]
 so that the range $-\infty<t<\infty$ corresponds to $0<\tau<\pi$. 
 
 The metric $\hg$ 
 is referred to in this context as \emph{the Einstein static cylinder}. We can add boundaries to the de Sitter space to compactify it in the Einstein cylinder: a past boundary at $\tau=0$ and a future boundary at $\tau=\pi$. 

 \item {\bf{anti-de Sitter space}}: One can similarly consider the hyperboloid
 \[T^2+W^2-X^2-Y^2-Z^2=H^{-2}=\mbox{  constant}\]
 in the 5-dimensional space with metric
 \[g=dT^2+dW^2-dX^2-dY^2-dZ^2.\]
 This is not simply-connected as it contains circles with $X,Y,Z$ constant and
 \[T^2+W^2=\mbox{ constant}\]
which are time-like (and therefore a causal problem) but the universal cover is free of these pathologies and is the space of constant curvature known as anti-de Sitter space. It's an exercise to introduce coordinates 
(start with $T+iW=H^{-1}e^{i\tau}\cosh(HR)$) and find the metric as
\[g=H^{-2}\cosh^2(HR)d\tau^2-(dR^2+H^{-2}\sinh^2(HR)(d\theta^2+\sin^2\theta d\phi^2)),\]
  and then calculate
 \[C_{abc}^{\;\;\;\;\;d}=0,\;\;\;R_{ab}=-3H^2g_{ab}.\]
 This is a warped product metric on $\mathbb{R}\times\mathbb{H}^3$. If we rescale with $\Omega=H\mbox{sech}(HR)$ and introduce the new radial coordinate $r$ by
 \[\sin r=\tanh(HR)\]
 then
 \[\hg=d\tau^2-(dr^2+\sin^2r(d\theta^2+\sin^2\theta d\phi^2)),\]
 which is (\ref{esc}) again, but the range $0\leq R<\infty$ of the anti-de Sitter radial coordinate corresponds to $0\leq r<\pi/2$ in the $\mathbb{S}^3$ of the Einstein static cylinder. 
 Thus anti-de Sitter space is also conformally related to a piece of the Einstein static cylinder consisting of the product 
 of the time-axis with a ball of finite radius in the $\mathbb{S}^3$ factor. The boundary $r=\pi/2$ is a time-like surface with the product metric on $\mathbb{R}\times\mathbb{S}^2$.
 \item {\bf{Minkowski space}}: Taking this in spherical polars for the space part, the metric is
 \[g=dT^2-(dR^2+R^2(d\theta^2+\sin^2\theta d\phi^2)).\]
 Introduce null coordinates
 \[u=(T-R)/2,\;\;v=(T+R)/2\]
 to obtain
 \[g=4dudv-(v-u)^2(d\theta^2+\sin^2\theta d\phi^2),\]
 and then set $u=\tan p,v=\tan q$ and $\Omega=\cos p\cos q$ to obtain
 \[\hg=4dpdq-\sin^2(q-p)(d\theta^2+\sin^2\theta d\phi^2).\]
 Now put $p=(\tau-r)/2,q=(\tau+r)/2$ to obtain (\ref{esc}) again.
 
 The region this time is a triangle in the $(p,q)$-plane
 \[-\frac{\pi}{2}<p\leq q<\frac{\pi}{2}\]
 corresponding to the triangle
 \[-\pi<\tau-r<\pi,\;\;-\pi<\tau+r<\pi,\;\;r\geq 0.\]
 The boundary of Minkowski space in the Einstein static cylinder consists of the past light cone (conventionally called $\scri^+$ and pronounced `scri-plus') of the point $i^+=\{\tau=\pi,r=0\}$ and the future 
 light cone ($\scri^-$ or `scri-minus') of 
 the point $i^-=\{\tau=-\pi,r=0\}$, which meet at the point $i^0=\{\tau=0,r=\pi\}$. These parts of the boundary have conventional names, motivated by the classes of geodesics which have them as end-points: 
 $i^+$ is \emph{future time-like infinity}, $i^-$ is \emph{past time-like infinity}, 
 $i^0$ is \emph{space-like infinity}, $\scri^+$ is \emph{future null infinity} and $\scri^-$ is \emph{past null infinity}.
\end{itemize}
\subsubsection{Asymptotic simplicity}\label{SSS1.2.4}
In the early days of GR, one understood an isolated system to be one which could be expressed in coordinates such that the metric approached flat space at a suitable rate in a suitable radial coordinate. A geometrical 
and coordinate-independent notion capturing this idea was introduced by Penrose in the early 1960's, motivated by the examples above. The definition \cite{pr} can be given as follows:

{\bf{Definition}}: A smooth space-time $M$ with metric $g$ is \emph{asymptotically simple} if there is a smooth manifold $\hat{M}$ with boundary $\scri$ and metric $\hg$ and a smooth scalar function $\Omega$ such that
\begin{enumerate}
 \item $M=\mbox{int }\hat{M}$,
 \item $\hg_{ab}=\Omega^2g_{ab}$ in $M$,
 \item $\Omega>0$ in $M$; $\Omega=0,\nabla_a\Omega\neq 0$ on $\scri$,
 \item every null geodesic in $M$ has a future and a past endpoint on $\scri$. 
\end{enumerate}
One can modify the definition to have lower degrees of differentiability, defining e.g. \emph{$C^k$-asymptotic simplicity}. 

\medskip

Condition 4 is included to ensure that one has all of $\scri$ and to exclude trivial examples with empty $\scri$, but would be too strong for example in space-times with black holes. Thus one introduces \emph{weak asymptotic simplicity}: a space-time $M$ 
is weakly asymptotically simple (or WAS) if there exists an asymptotically-simple $M'$ and a neighbourhood $O'$ of $\scri$ in $\hat{M}'$ such that $O'\cap M'$ is isometric to a subset of $M$.

\medskip

We draw some conclusions from the assumption that $M$ is WAS:
\begin{itemize}
 \item From (\ref{c3}), replacing $\Omega$ by its inverse:
\[\Omega^{-2}R=\hat{R}+6\Omega\hat{\Box}(\Omega^{-1})=\hat{R}-6\Omega^{-1}\hat{\Box}\Omega+12\Omega^{-2}\hg^{ab}\Omega_a\Omega_b,\]
and so
\be\label{c6}R=\Omega^2\hat{R}-6\Omega\hat{\Box}\Omega+12\hg^{ab}\Omega_a\Omega_b.\ee
Now at $\scri$, $\Omega$ vanishes, $\Omega_a$ is the tangent to $\scri$, nonvanishing by assumption, and we can assume $\hat{R}$ and $\hat{\Box}\Omega$ are finite. The Einstein equations with cosmological constant $\lambda$ imply
\[R=4\lambda+8\pi GT,\]
where $T$ is the trace of the energy monmentum tensor. We conclude by evaluating (\ref{c6}) at $\scri$ that, provided $T$ is zero in a neighbourhood 
of $\scri$, \emph{the surface $\scri$ is space-like, time-like or null (i.e. its normal is time-like, space-like or null)  
according as $\lambda$ is positive, negative or zero}. Note that this is in line with the explicit examples of de Sitter space, anti-de Sitter space and Minkowski space treated earlier.
\item From (\ref{c1}) replacing $\Omega$ by its inverse:
\[\Omega R_{ab}=\Omega\hat{R}_{ab}-2\hat{\nabla}_a\Omega_b-\hg_{ab}\hg^{cd}(\hat{\nabla}_c\Omega_d-3\Omega^{-1}\Omega_c\Omega_d).\]
Take the trace-free part to deduce that \emph{if the trace-free part of $T_{ab}$ is zero (or just bounded) in a neighbourhood of $\scri$ then the trace-free part of $\hat{\nabla}_a\Omega_b$ vanishes at $\scri$}. 
This means that $\scri$ is umbilic if time-like or space-like and shear-free if null. Refining the choice of $\Omega$ allows w.l.o.g. the assumption that $\scri$ is actually extrinsically flat in the time-like or space-like cases and 
expansion-free in the null case.
\item Using (\ref{c1}) and (\ref{2.2}) we calculate
\[\hat{\nabla}_d\hat{C}_{abc}^{\;\;\;\;\;d}=\nabla_dC_{abc}^{\;\;\;\;\;d}+C_{abc}^{\;\;\;\;\;d}\Upsilon_d.\]
Multiply by $\Omega$ and take the limit at $\scri$. From the Bianchi identity (\ref{9}) and the Einstein field equations, 
if the space-time is vacuum near $\scri$ or if the matter content (and therefore the physical Ricci curvature) falls off fast enough then the term $\nabla_dC_{abc}^{\;\;\;\;\;d}$ 
goes to zero at $\scri$ while the term on the left is bounded: we conclude that $\hat{C}_{abc}^{\;\;\;\;\;d}\Omega_d$ vanishes at $\scri$. In the time-like or space-like case this is sufficient to conclude that 
$\hat{C}_{abc}^{\;\;\;\;\;d}=0$ at $\scri$. In the null case the conclusion follows if one can show that each component of $\scri$ is topologically $\mathbb{R}\times\mathbb{S}^2$ 
(which is possible but intricate; see \cite{p2},\cite{n1}).

\end{itemize}

\subsubsection{Conformal geodesics}\label{SSS1.2.5}
These are a class of curves with better properties under conformal rescaling than metric geodesics, (\cite{be},\cite{fs},\cite{sch}). There are two slightly different ways to describe them:
first, a conformal geodesic is a curve $\gamma$ with
tangent vector $v^a$ and a one-form $b_a$ given along it and
satisfying the system:
\begin{eqnarray} 
v^c\nabla_cv^a&=&-2(v^cb_c)v^a+g^{ac}b_c(g_{ef}v^ev^f),  \label{cg1}\\
v^c\nabla_cb_a&=&(v^cb_c)b_a-\frac12g_{ac}v^c(g^{ef}b_eb_f)+L_{ac}v^c,
\label{cg2}
 \end{eqnarray}
where $L_{ab}$ as in (\ref{8}). In this form, the system transforms to itself under conformal rescaling with

\begin{eqnarray} \hat{v}^a&=&v^a,\label{cs2}\\
\hat{b}_a&=&b_a-\Upsilon_a.\label{cs3} 
\end{eqnarray}
It's clear from (\ref{cg1}) that a null conformal geodesic is in fact a null metric geodesic and, conversely, given a null metric geodesic one can find $(v^a,b_a)$ to make it a conformal geodesic. 
In \cite{sch} an interpretation of non-null conformal geodesics was given as follows: given a segment of a non-null conformal 
geodesic on which $(v^a,b_a)$ are finite and nonzero solve for $\Omega$ in
\[v^a\Omega_a=\Omega v^ab_a;\]
this gives $\Omega$ along the segment; find it in a neighbourhood of the segment such that $\Upsilon_a=b_a$ at the segment and
\[\nabla_a\Upsilon_b=b_ab_b-\frac12g_{ab}b^cb_c+L_{ab},  \]
again, along the segment. There will be many such $\Omega$ but now rescaling by one such reduces $b_a$ to zero by (\ref{cs3}) and $L_{ab}$ to zero at the segment by (\ref{l4}). Now by (\ref{cg1}) the conformal geodesic is 
a metric geodesic so

\medskip

\emph{a conformal geodesic $\gamma$ is a metric geodesic in a rescaled metric for which the Ricci tensor vanishes at $\gamma$.}

\medskip

A conformal geodesic admits a preferred parameter $\sigma$ defined up
to choice of origin by
\be\label{cg3}v^c\nabla_c\sigma=1.\ee
There is a reparametrisation freedom of M\"obius transformations in $\sigma$:
\be\label{cg4} \sigma\rightarrow\tilde\sigma=\frac{a\sigma+b}{c\sigma+d},\ee
with $ad-bc=1$, so that $\sigma$ may be called a \emph{projective parameter}, provided this is accompanied by a transformation of $v^a$ and $b_a$:
\begin{eqnarray}\tilde{v}^a&=&(c\sigma+d)^{2}v^a\label{cg5}\\
\tilde{b}_a&=&b_a+fg_{ac}v^c\label{cg6}\end{eqnarray}
where
\[f=-2(g_{ef}v^ev^f)^{-1}c(c\sigma+d)^{-1}.\]
This transformation draws attention to a problem in working with conformal geodesics: it is possible for $\tilde{v}^a$
to vanish and $\tilde{b}_a$ to be singular at a regular point both of the manifold and of the curve, by choice of
projective parameter. 

One way to avoid this problem is to introduce a \emph{third-order form} of the equations.  How this is done depends slightly on the signature so we'll restrict to a time-like 
conformal geodesic in a Lorentzian space-time. (See \cite{l2} for a discussion and application of null conformal geodesics.)

Introduce the unit tangent $u^a=\chi^{-2}v^a$, where $\chi^4=g_{ef}v^ev^f$. Now the curve is parametrised by proper-time and (\ref{cg1}) can be solved for $b_c$ in
terms of the acceleration $a^c=u^a\nabla_au^c$ as
\be\label{3d}b^c=a^c+(u^ab_a)u^c,\ee
where $u^ab_a=-2\chi^{-1}u^a\nabla_a\chi=-2\dot{\chi}/\chi$ and the overdot is differentiation w.r.t. proper-time.
Use this in (\ref{cg2})
to find
\be\label{3e}u^c\nabla_ca^b=u^b(-g_{ef}a^ea^f-L_{ef}u^eu^f)+L^b_cu^c,\ee
and
\be\label{3f}\ddot{\chi}=-\frac14(g_{ef}a^ea^f+2L_{ef}u^eu^f)\chi.\ee
In this form of the equations, (\ref{3e}) is a third-order equation for the curve in proper time and then (\ref{3f}) is a second-order equation for $\chi$ which with the acceleration $a^b$ determines the one-form $b_c$. 
To obtain the projective parameter $\sigma$ from proper-time note that
\[1=v^c\nabla_c\sigma=\chi^2u^c\nabla_c\tau=\chi^2\frac{d\sigma}{dt},\]
so that $d\sigma=dt/\chi^2$. If we choose two solutions $\chi_1,\chi_2$ of (\ref{3f}) with unit Wronskian:
\[\chi_1\dot{\chi}_2-\chi_2\dot{\chi}_1=1,\mbox{   so that   }\frac{d}{dt}\left(\frac{\chi_2}{\chi_1}\right)=\frac{1}{\chi_1^2}\mbox{  and  }\chi_1^{-2}dt=d\left(\frac{\chi_2}{\chi_1}\right),\]
then an allowed choice for $\sigma$ is $\chi_2/\chi_1$. Different choices of the solutions $\chi_i$ determine projective transformations of $\sigma$, and zeroes and poles of the projective parameter are tied to zeroes and poles of the solutions 
of (\ref{3f}).

\subsubsection{The Weyl connection associated with a conformal geodesic}
\label{SSS1.2.6}
Given a conformal geodesic $\gamma$, one may define a conformally-invariant propagation of vectors along it according to
\be\label{3g} v^bD_be^a:=v^b\nabla_be^a+(v^cb_c)e^a+(e^cb_c)v^a-g^{ac}b_c(g_{ef}v^ev^f)=0.\ee 
From the way this is written, it is evidently parallel propagation in a connection which differs from the metric connection by a change of Christoffel symbols:
\[\Gamma\rightarrow\tilde{\Gamma}=\Gamma+\delta^a_bb_c+\delta^a_cb_b-g^{ad}b_dg_{bc}.\]
This connection, $D_a$, is symmetric and preserves the metric up to scale:
\[D_ag_{bc}=-2b_ag_{bc},\]
so it's a \emph{Weyl connection}. In particular $v^a$ is propagated in this connection, since (\ref{cg1}) is just
\[v^bD_bv^a=0,\]
and the statement
\[v^bD_be^a=0\]is conformally invariant (i.e. preserved with the rescaling $\hat{e}^a=e^a$). We can define an orthogonal basis of vectors $e^a_\alpha$ with dual basis $\theta_a^\alpha$ along $\gamma$ and then the components
\[C_{\alpha\beta\gamma}^{\;\;\;\;\;\delta}:=C_{abc}^{\;\;\;\;\;d}e^a_\alpha e^b_\beta e^c_\gamma\theta_d^\delta\]
of the Weyl tensor are conformally-invariant, as are components of derivatives $D_{a_1}\cdots D_{a_n}C_{abc}^{\;\;\;\;\;d}$. This will be relevant in section \ref{SSS3.2.4}.

\subsubsection{Some general comments on geodesics and conformal geodesics}
\label{SSS1.2.7}
\begin{itemize}
\item In 4 dimensions, there is a 9-parameter family of unparametrised conformal geodesics and a 5-parameter family of unparametrised metric geodesics (so there are clearly `more' conformal geodesics).
\item Metric geodesics are not conformal geodesics in general but will be in a Ricci-flat space, or if the velocity vector is an eigenvector of the Ricci tensor (an example of this case is provided by 
the matter flow lines in a perfect-fluid FRW cosmological model). Null geodesics however are always conformal geodesics.
 \item In Euclidian space, the conformal geodesics are planar circles (or straight lines). In Minkowski space-time the constant acceleration world-lines are conformal geodesics -- they go through $\scri$ 
 and so components in a constant basis of the velocity $u^a$ become infinite. (See \cite{t1} for more examples and results).
 \item In general, neither Killing vectors nor conformal Killing vectors give conserved quantities for conformal geodesics, so the equations have been hard to integrate explicitly (again, there are examples 
 when it can be done in \cite{t1}).
 \item In the Einstein static cylinder, the vector field $\partial/\partial\tau$ is tangent to conformal geodesics and to metric geodesics. A choice of projective parameter is $\sigma=\tan\tau$ so that $\sigma$ 
 has an infinite sequence of poles and zeroes for an infinite range in $\tau$. If we consider an FRW metric with $\mathbb{S}^3$ space sections in the form:
 \[g=dt^2-t^{2k}(dr^2+\sin^2r(d\theta^2+\sin^2\theta d\phi^2)),\]
 then with conformal factor $\Omega=t^{-k}$ this rescales to the Einstein static cylinder
 \[\hg=t^{-2k}(dt^2-t^{2k}(dr^2+\sin^2r(d\theta^2+\sin^2\theta d\phi^2))\]\[=d\tau^2-(dr^2+\sin^2r(d\theta^2+\sin^2\theta d\phi^2))\]
 with $\tau=\pm t^{1-k}/(1-k)$. Now with $k>1$ the initial singularity of the FRW metric at $t=0$ is sent to $\tau=\pm\infty$ and in particular it hasn't been added as a boundary. This can be seen to have happened because there are 
 infinitely many zeroes and poles of the projective parameter along the conformal geodesic which is a matter flow line in the FRW metric. Thus it makes sense to say that for $k>1$ the initial singularity was 
 `conformally infinitely far away'. This will be significant in an example below.
\end{itemize}

\subsection{Behaviour of matter models under conformal rescaling}\label{SS1.3}
\subsubsection{Source-free Maxwell fields}\label{SSS1.3.1}
The Maxwell field can be regarded as a 2-form ${\bf{F}}$ and then the source-free Maxwell equations require this and its dual to be closed:
\[d{\bf{F}}=0=d{\bf{F}}^*.\]If we assume that under conformal rescaling ${\bf{F}}\rightarrow{\bf{\hat{F}}}={\bf{F}}$ (i.e. conformal weight 0) then also ${\bf{F}}^*\rightarrow{\bf{\hat{F}}}*={\bf{F}}^*$ by (\ref{d1}) so that both Maxwell 
equations are preserved. In index notation
\[\hat{F}_{ab}=F_{ab},\;\;\hat{F}^*_{ab}=F^*_{ab}=\frac12\epsilon_{ab}^{\;\;\;\;cd}F_{cd}.\]
For the energy-momentum tensor we have
\[T_{ab}=-\frac{1}{4\pi}\left(F_{ac}F_b^{\;\;c}-\frac14F_{cd}F^{cd}g_{ab}\right),\]
and it's easy to see that
\be\label{d2}\hat{T}_{ab}=\Omega^{-2}T_{ab},\ee
and that this preserves the conservation equation. 

\medskip 
Note
\begin{itemize}
 \item for \emph{any} trace-free $T_{ab}$ the transformation (\ref{d2}) preserves the conservation equation, but if $T_{ab}$ is not trace-free then no simple 
transformation preserves conservation;
\item while any Killing vector $K^a$ generates a conserved current $J_a:=T_{ab}K^b$ from any conserved $T_{ab}$, a \emph{conformal} Killing vector $X^a$ will produce a 
conserved current $\tilde{J}_b:=T_{ab}X^b$ from a conserved \emph{trace-free} $T_{ab}$, but not otherwise. We'll see an application of this in section \ref{SSS3.6.2}.
\end{itemize}

\subsubsection{Scalar fields}\label{SSS1.3.2}
The Klein-Gordon or massless scalar field equations do not have a simple transformation but there is a modification which does. Following the presentation in \cite{l1}, consider the tensor
\[D_{ab}[\phi,g_{cd},\alpha]:=4\phi_a\phi_b-g_{ab}g^{cd}\phi_c\phi_d-2\phi\nabla_a\phi_b+2\phi^2L_{ab}+2\alpha\phi^4g_{ab}\]
with
\[L_{ab}=-\frac12R_{ab}+\frac{1}{12}Rg_{ab}\]
as usual, $\alpha$ a real constant and $\phi$ a real scalar field. Define
\[Q(\phi):=\Box\phi+\frac16R\phi-4\alpha\phi^3.\]
We claim that, if $\hat{\phi}=\Omega^{-1}\phi$ (so $\phi$ has conformal weight -1) then 
\[D_{ab}[\hat{\phi},\hg_{cd},\alpha]=\Omega^{-2}D_{ab}[\phi,g_{cd},\alpha],\]
(this is just a matter of checking). Note that
\[g^{ab}D_{ab}=-2\phi Q(\phi),\]
and that 
\[\nabla^aD_{ab}=4Q\phi_b-2\phi Q_b.\]
Therefore if we require $\phi$ to satisfy the field equation $Q=0$ i.e.
\be\label{d4}
\Box\phi+\frac16R\phi-4\alpha\phi^3=0\ee
which is a \emph{non-minimally coupled Klein-Gordon equation}, then $D_{ab}[\phi,g_{cd},\alpha]$ is trace-free and divergence-free and by the argument around (\ref{d2}) so is $D_{ab}[\hat{\phi},\hg_{cd},\alpha]$ in 
the rescaled connection. Therefore $\hat\phi$ satisfies (\ref{d4}) w.r.t. $\hg$ (which can be checked directly). In this sense the field equation (\ref{d4}) is conformally invariant.

A scalar field satisfying (\ref{d4}) is conveniently called a \emph{conformal scalar} and $D_{ab}[\phi,g_{cd},\alpha]$ can be taken as its energy momentum tensor. We'll come back to this.

\subsubsection{The Vlasov and Boltzmann equations}\label{SSS1.3.3}
In statistical mechanics a distribution of matter is defined by its distribution function, a non-negative function $f(q^a,p_a)$ on the phase space $T^*(M)$. For simplicity, the particles are often thought of as belonging to 
a single species and the support of $f$ is confined to the future mass shell $\{g^{ab}p_ap_b=m^2,\;\;p_0>0\}$. This matter model typically does not have good behaviour under conformal rescaling (unsurprisingly) but it's a different 
story if one restricts the support to the future null cone $N^+_q$ at the point labelled $q$ --  this is \emph{massless Vlasov} or \emph{massless Boltzmann}. 

Vlasov is the case of collisionless matter and then $f$ is constant along the geodesic flow:
\[\{g^{ab}p_ap_b,f(q,p)\}=0\]
equivalently
\be\label{v2}\mathcal{L}f:=g^{ab}p_a\frac{\partial f}{\partial q^b}-p_ap_b\frac{\partial g_{ab}}{\partial q^c}\frac{\partial f}{\partial p_c}=0, \ee
which can conveniently be called the Vlasov equation.

\medskip

For the Boltzmann case one needs a collision-term on the right-hand-side of the Vlasov equation and this is typically quadratic in $f$. 

\medskip

In either case, from $f$ one defines the energy-momentum tensor by
\be\label{v1}
T_{ab}=\int_{N_q} p_ap_bf(q,p)\omega_p=\frac{1}{\sqrt{-g}}\int_{N_q} p_ap_bf(q,p)\frac{d^3p}{p^0},\ee
where $\omega_p=d^3p/(\sqrt{-g}p^0)$ is a Lorentz-invariant volume form on $N_q$ which we'll explain below, and then this can be used in the Einstein field equations. It is clear that $T_{ab}$ in (\ref{v1}) is trace-free and 
it's an exercise to check that it 
is divergence-free by virtue of (\ref{v2}) (this is much easier to see in inertal coordinates).

To find the behaviour under conformal rescaling we first note that $\omega_p$ can be written
\[\omega_p=\frac{\epsilon^{abcd}T_adp_b\wedge dp_c\wedge dp_d}{6g^{ef}T_ep_f}=\frac{d^3p}{\sqrt{-g}p^0}\]
where $T^e$ is an arbitrary time-like vector. This evidently picks up a factor $\Omega^{-2}$ under rescaling if we assume $\hat{p}_a=p_a$, so with $\hat{f}=f$ we'll have $\hat{T}_{ab}=\Omega^{-2}T_{ab}$ which is the 
correct rescaling to preserve the conservation equation.

\medskip

For Boltzmann we need to know more about $C(f,f)$. As is standard, we assume \emph{binary collisions} so that a pair of particles with null 4-momenta $p^a,q^a$ collide to produce a pair with null 
4-momenta $p'^a,q'^a$ (or vice versa) and 
we assume conservation of 
4-momentum so that the total 4-momentum in the collision is
\[P^a:=p^a+q^a=p'^a+q'^a.\]
The Boltzmann equation is
\[{\mathcal{L}}f=C(f,f),\]
where the R.H.S., called the \emph{collision term} is an integral
\[C(f,f)=\int(f(t,p')f(t,q')-f(t,p)f(t,q))k(s,\theta)\omega_q\xi_{p'}\]
over allowed $q,p',q'$. Here $k(s,\theta)$ is a function to be specified and (frequently but not always) known as the scattering-cross-section, the scalar $s$ is defined as
\[s=g^{ab}P_aP_b=2g^{ab}p_aq_b
,\]
$\theta$ is the usual scattering angle, determined from
\[g^{ab}p_ap'_b=\frac{s}{4}(1-\cos\theta),\]
and $\omega_q$ is the 3-form introduced above, on the null-cone for $q_a$. The integral is over allowed $q$, which is the 3-dimensional null cone for $q$, and allowed $p',q'$ constrained by $p'+q'=p+q=P$ say. There is a 2-dimensional 
allowed set of $p'$, since
\[g^{ab}p'_ap'_b=0=g^{ab}P_a(P_b-2p'_b),\]
and given $p'$ and $q$ there is no freedom in $q'$. Therefore $\xi_{p'}$ is a 2-form on the space of allowed $p'_a$ (called a \emph{Leray form} in the French literature, \cite{ycb}), and it can be defined as
\[\xi_{p'}=-\frac{\epsilon^{abcd}p'_aq'_bdp'_c\wedge dp'_d}{2(g^{kl}p_kq_l)^2}.\]
This makes manifest the conformal properties of $\xi_p$: it is in fact unchanged by conformal rescaling.

\medskip 

(The definition of $\xi_p$ can be motivated as follows: on the submanifold $\Sigma_{pqP}$ of  $N_p\times N_q$, the product of null-cones in $p$ and in $q$, on which $p_a+q_a=P_a$ for fixed $P_a$ we want to 
define a 2-form $\xi$ so that, on  $N_p\times N_q$,
\[\xi\wedge(\frac{1}{24}\epsilon^{abcd}dP_a\wedge dP_b\wedge dP_c\wedge dP_d)=\omega_p\wedge\omega_q.\]

This is because it ensures that
\[\int_{N_p\times N_q} F\delta^{(4)}(P_a-p'_a-q'_a)\omega_{p'}\wedge\omega_{q'}=\int_{\Sigma_{p'q'T}}F\xi_{p'}\]
with $F$ suitably restricted, and the $\delta$-function included to enforce conservation of 4-momentum.)
%
%
%
%

If we now have an explicit expression for $k(s,\theta)$ then we will know the transformation of the Boltzmann equation. 
\subsubsection{The `conformal-to-Einstein' condition}\label{SSS1.3.4}
While not actually a matter equation of motion, it is interesting to consider the equation (studied in \cite{leb} and the work of Friedrich e.g. \cite{f1}):
\be\label{v3}\nabla_a\nabla_b\sigma+\sigma L_{ab}+\rho g_{ab}=0.\ee
This is equivalent to the statement that the trace-free part of $\nabla_a\nabla_b\sigma+\sigma L_{ab}$ is zero, since $\rho$ can be eliminated in terms of $\sigma$ and $\Box\sigma$:
\[\rho=-\frac14(\Box\sigma-\frac16R\sigma).\] Equation (\ref{v3}) has a significance that 
we'll come to but the first observation is that it is conformally-invariant if we accompany (\ref{2.1}) with
\[\sigma\rightarrow\hat{\sigma}=\Omega\sigma,\]
that is, $\sigma$ has conformal weight 1. This follows rapidly from (\ref{l4}), and we also obtain 
\[\rho\rightarrow\hat{\rho}=\Omega^{-1}(\rho-\Upsilon^c\sigma_c-\frac12\sigma\Upsilon^c\Upsilon_c).\]
Now suppose that one has a solution of (\ref{v3}), then in regions in which it has no zeroes we may use $\Omega=\sigma^{-1}$ as a conformal factor to obtain $\hat\sigma=1$, when 
$\hat{\nabla}_a\hat{\nabla}_b\hat{\sigma}=0$ and so 
the trace-free part of $\hat{L}_{ab}$ is zero. Therefore a space-time admits nontrivial solutions of (\ref{v3}) if and only if it is locally conformal to an Einstein space, \cite{leb}. 
\subsubsection{Two, more complicated examples }\label{SSS1.3.5}
We'll review two, more complicated examples, both of which have actually arisen in cosmological studies.

\medskip
\begin{itemize}
 \item{\bf{Massless viscous magnetohydrodynamics (MHD)}}

Following \cite{sb} we note that the equations for this are conformally-invariant. The model is radiation fluid which moves through a magnetic field which it itself generates. Suppose the fluid has unit velocity $u^a$, 
density $\rho$ and pressure $p=\rho/3$. There is a standard kinematic decomposition of the covariant derivative of the velocity:
\[\nabla_au_b=u_aA_b+\omega_{ab}+\sigma_{ab}+\frac13\theta h_{ab}\]
with $h_{ab}=g_{ab}-u_au_b$, which is the projection orthogonal to the fluid flow, and $A_b,\omega_{ab},\sigma_{ab},\theta$ are respectively the acceleration, twist, shear and expansion of the fluid flow. These are fixed by 
requiring
\[u^aA_a=0=\omega_{ab}u^b=\sigma_{ab}u^b=\omega_{(ab)}=\sigma_{[ab]}=h^{ab}\sigma_{ab}.\]
From the discussion in \cite{wei}, we'll suppose that there is \emph{shear viscosity} but no \emph{bulk viscosity}. This is expressed by adding a term $-\eta\sigma_{ab}$, with $\eta$ the coefficient of shear viscosity, 
to the fluid energy-momentum tensor so that this becomes
\be\label{mh1}T_{ab}^F=\frac13\rho(4u_au_b-g_{ab})-\eta\sigma_{ab}.\ee
In what follows, we allow $\eta$ to be a function of position.

There is a Maxwell field $F_{ab}$ which is generated by a current density $J_a$ from Ohm's Law
\be\label{mh2}J_a-u_a(u^bJ_b)=\sigma F_{ab}u^b\ee
where $\sigma$ is conductivity, again allowed to be a function of position, so that
\[\nabla_{[a}F_{bc]}=0,\;\;\;\nabla_aF^{ab}=4\pi J^b.\]
The presently undefined component $J_au^a$ is the charge density and we'll suppose this is zero. The Maxwell contribution to the energy-momentum tensor is
\[T_{ab}^M=-\frac{1}{4\pi}\left(F_{ac}F_b^{\;\;c}-\frac14F_{cd}F^{cd}g_{ab}\right),\]
and the total energy-momentum tensor is the sum of this and the fluid energy-momentum tensor \ref{mh1}. Both energy-momentum tensors are separately trace-free, which is an indicator of good conformal behaviour.

Under rescaling
\[g_{ab}\rightarrow\hg_{ab}=\Omega^2g_{ab},\]
we'll have $u^a\rightarrow\hat{u}^a=\Omega^{-1}u^a$, whence the kinematic quantities must transform as
\[\hat{A}_b=A_b-(\Upsilon_b-u_b(u^c\Upsilon_c)),\;\;\hat{\omega}_{ab}=\Omega\omega_{ab},\;\;\hat{\sigma}_{ab}=\Omega\sigma_{ab},\;\;\hat{\theta}=\Omega^{-1}\theta.\]
From the remarks in Section \ref{SSS1.3.1} we want the energy-momentum tensors to have conformal weight $-2$ and this is achieved with
\be\label{mh3}\hat{\rho}=\Omega^{-4}\rho,\;\;\;\hat{\eta}=\Omega^{-3}\eta,\ee
and Ohm's Law transforms properly with
\be\label{mh4}\hat{J}_a=\Omega^{-2}J_a,\;\;\;\hat{\sigma}=\Omega^{-1}\sigma.\ee
Note once again that these rescalings require, for consistency, that $\sigma$ and $\eta$ are functions of position i.e. are not constant.

\medskip

The conformal invariance of this system, pointed out by \cite{sb}, was used by them in an FRW cosmology to solve curved-space MHD from flat-space MHD.
\item{\bf{Massless Vlasov with Yang-Mills}}

Here we follow \cite{ycb} and \cite{t4}. The model has massless Vlasov matter carrying a Yang-Mills charge. Quantities will typically carry Yang-Mills indices which we'll take to be Greek. Thus there is Yang-Mills potential $A^\alpha_a$ 
which gives rise to a Yang-Mills field $F_{ab}^\alpha$ according to
\be\label{ym1}F_{ab}^\alpha=\nabla_aA_b^\alpha-\nabla_bA_a^\alpha +c^\alpha_{\;\;\beta\gamma}A_a^\beta A_b^\gamma,\ee
where $c^\alpha_{\;\;\beta\gamma}$ are the structure constants of whichever Yang-Mills group $\mathcal{G}$, conveniently assumed to be semi-simple, has been chosen. To describe the matter there is a distribution function 
$f(q^a,p_a,Q^\alpha)$ supported on future null-cones $N^+_q=\{g^{ab}(q)p_ap_b=0, p_0>0\}$ 
and dependent on a Yang-Mills charge vector $Q^\alpha$. This defines an energy-momentum tensor as in (\ref{v1}) and also a current vector 
\be\label{ym2}T_{ab}^V=\int_{N_q} p_ap_bf(q,p)\omega_p,\;\;J_a^\alpha=\int\int Q^\alpha p_af\omega_p\omega_Q,\ee
where $\omega_Q$ is a volume form on the space of $Q^\alpha$.

\medskip

The distribution function satisfies a Vlasov equation
\be\label{ym3}\frac{\partial f}{\partial q^a}\frac{dq^a}{ds}+\frac{\partial f}{\partial p_a}\frac{dp_a}{ds}+\frac{\partial f}{\partial Q^\alpha}\frac{dQ^\alpha}{ds}=0,\ee
and the particles follow a Lorentz force-law adapted for Yang-Mills theory:
\begin{eqnarray}\label{ym4}
 \frac{dq^a}{ds}&=&g^{ab}p_b\\
 \frac{dp_a}{ds}&=&-\frac12g^{bc}_{\;\;\;,a}p_bp_c+Q^\alpha\eta_{\alpha\beta}F^\beta_{ab}g^{bc}p_c\label{ym5}\\
 \frac{dQ^\alpha}{ds}&=&-c^\alpha_{\;\;\beta\gamma}A^\beta_bg^{bc}p_cQ_\gamma\label{ym6}. 
\end{eqnarray}
Here $\eta_{\alpha\beta}$ is the metric on $\mathcal{G}$, which exists by assumption, and the metric $g^{bc}$ has been written inexplicitly to help with the book-keeping.

The remaining Yang-Mills equation is
\be\label{ym7}\nabla_aF^{ab\alpha}+c^\alpha_{\;\;\beta\gamma}A^\beta_aF^{ab\gamma}=4\pi J^{b\alpha},\ee
and the Yang-Mills contribution to the energy-momentum tensor is
\be\label{ym8}T^{YM}_{ab}=-\eta_{\alpha\beta}(F_{ac}^{\;\;\;\alpha} F_b^{\;\;c\,\beta}-\frac14F_{cd}^\alpha F^{cd\,\beta}g_{ab}).\ee

\medskip

Under conformal rescaling, following the example of Sections \ref{SSS1.3.1} and \ref{SSS1.3.3}, $f, p_a, Q^\alpha$ and $A_a^\alpha$ are unchanged, therefore so is $F_{ab}^\alpha$, while $T_{ab}^V,J_a^\alpha$ and $T^{YM}_{ab}$ all have 
weight $-2$. Proper time in the Lorentz force-law and the Vlasov equation will change according to $d/d\hat{s}=\Omega^{-2}d/ds$, while the right-hand-sides in the Lorentz force-law 
all have weight $-2$ (this relies on $g^{cd}p_cp_d=0$) so these equations are invariant. 
Finally one readily checks that (\ref{ym7}) is invariant ($F^{ab\alpha}$ and $J^{b\alpha}$ both have weight $-4$ and the terms in $\Upsilon$ from the derivative cancel each other).

This system has been considered as a source for the Einstein equations: local existence was proved in \cite{ycb}, and conformal gauge singularities (looking ahead to Section \ref{SS3.2}) were investigated in \cite{t4}.
\end{itemize}

\section{Cosmology}\label{S2}
As a preamble to CCC, we'll review some elementary mathematical cosmology with particular interest in the case of positive $\lambda$.
\subsection{Friedmann-Lemaitre-Robertson-Walker (FLRW) models}\label{SS2.1}
For these the metric can be written
\[g=dt^2-(a(t))^2d\sigma^2_k\]
where the spatial metric is
\[d\sigma^2_k=dr^2+(f_k(r))^2(d\theta^2+\sin^2\theta d\phi^2)\]
with the familiar choices
\[f_1=\sin r,\;\;f_0=r,\;\;f_{-1}=\sinh r.\]
The Ricci tensor $R_{ab}$ necessarily takes the form $A(t)u_au_b+B(t)g_{ab}$ with $u_a=\nabla_at$ so these are naturally adapted to the perfect fluid matter model with fluid velocity $u^a$, and 
with a cosmological constant. The energy-momentum tensor 
for the matter is
\[T_{ab}=(\rho+p)u_au_b-pg_{ab}\]
and the EFEs including $\lambda$ are
\begin{eqnarray*}
R_{00}: 3a^{-1}\ddot{a}&=&-4\pi G(\rho+3p)+\lambda\\
R_{ij}: a^{-1}\ddot{a}+2a^{-2}\dot{a}^2+2ka^{-2}&=&4\pi G(\rho-p)+\lambda
\end{eqnarray*}
Eliminate $\ddot{a}$ from these to arrive at the Friedmann equation
\be\label{f11}\dot{a}^2+k=\frac{8}{3}\pi G\rho a^2+\frac13\lambda a^2.\ee
Dot this and eliminate $\ddot{a}$ again to arrive at the conservation equation
\be\label{f2}\dot{\rho}a+3\dot{a}(\rho+p)=0.\ee
The complete set of EFEs is equivalent to (\ref{f11}, \ref{f2}), together with an equation of state $p=f(\rho)$. 

Easy cases to solve are the polytropic equations of state $p=(\gamma-1)\rho$ with $1\leq\gamma\leq 2$, and the usual understanding is that $\gamma=1,\;p=0$ is dust, $\gamma=4/3$ is radiation ($T^a_a=0$), 
and $\gamma=2, dp/d\rho=1$ is `stiff' matter (for which the speed of sound equals the speed of light). In these cases, 
the conservation equation integrates to give 
\be\label{f3}\rho a^{3\gamma}=3\mu/8\pi G=\mbox{constant}\ee
where $\mu$ is a constant of integration characterising the density, and the Friedmann equation becomes
\[\dot{a}^2+k=\mu a^{2-3\gamma}+\frac13\lambda a^2.\]
If solutions start with a Big Bang at $t=0$ then for small $t$
\be\label{c9}a\sim t^{2/3\gamma},\;\;\rho\sim t^{-2},\ee
so that the Ricci tensor is indeed singular at the Bang (since $\rho$ is a component of the Ricci tensor in a parallelly-propagated frame, and diverges at the Bang) while if $\lambda=3H^2$ then for large $t$
\be\label{c4}a=a_1e^{Ht}+a_{-1}e^{-Ht}+\mbox{h.o.},\;\;\rho\sim e^{-3\gamma Ht},\ee
with $a_{-1}$ fixed by $a_1$ and $k$.

An interesting special case is radiation when $\rho =3 a^{-4}\mu/8\pi G$ and
\[\dot{a}^2+k=\mu a^{-2}+\frac13\lambda a^2.\]
Introduce conformal time $\tau$ by $d\tau=dt/a$ and then
\be\label{c10}\left(\frac{da}{d\tau}\right)^2=\mu-ka^2+\frac13\lambda a^4,\ee
an equation that we shall revisit.

\subsubsection{Finding the scale-factor for a two-component universe with positive cosmological constant}\label{SSS2.1.1}
This will be helpful below, so suppose then that the universe is filled with a mixture of dust and radiation which don't interact. From (\ref{f2}) and the discussion below it the two components have respective densities
\be\label{tc1}\rho_m=\frac{A}{a^3},\;\;\rho_\gamma=\frac{B}{a^4},\ee
for constants of integration $A,B$. Here the dark matter and what is conventionally called the baryon contribution is in $\rho_m$ and photons and massless neutrinos are in $\rho_\gamma$. The Friedmann equation (\ref{f11}) becomes
\[\dot{a}^2=-k+\kappa a^2(\rho_m+\rho_\gamma)+H^2a^2,\]
where $\kappa=8\pi G/3$ and $H^2=\lambda/3$. The general consensus is that $k$ contributes very little to this equation -- it is swamped by the densities at early times and by the cosmological constant term at late times -- so we shall 
set it to zero. Then there is a conventional parametrisation of the densities by comparison with the $H$-term: introduce the quantities
\[\alpha=\frac{\kappa\rho_m}{H^2},\;\;\beta=\frac{\kappa\rho_\gamma}{H^2},\]
where these are calculated \emph{at the present time}. One can find these tabulated in the literature and commonly accepted values are $(\alpha,\beta)=(0.45,1.4\times 10^{-4})$ (e.g.\cite{pdg} or \cite{bau}) while the $e$-folding time 
$H^{-1}$ is $1.74\times 10^{10}$ years, which is surprisingly close to the accepted 
age of the universe which is $t_0=1.38\times 10^{10}$ years (so $H^{-1}=1.26t_0$).

Writing $a_0$ for the scale factor at the present time, one can solve for $A,B$ as
\[\kappa A=\alpha a_0^3H^2,\;\;\;\kappa B=\beta a_0^4H^2,\] and write the Friedmann equation as an equation for $S=a/a_0$ as
\be\label{tc2}
S^2\dot{S}^2=H^2(\beta+\alpha S+S^4).\ee
Recall $dS/d\tau=a\dot{S}=a_0S\dot{S}$, so we can integrate (\ref{tc2}) to obtain either $\tau$ or $t$ as
\be\label{tc3}
\int_0^S\frac{dS}{(\beta+\alpha S+S^4)^{1/2}}=a_0H\tau,\ee
or
\be\label{tc4}
\int_0^S\frac{SdS}{(\beta+\alpha S+S^4)^{1/2}}=Ht,\ee
where we've chosen the origins to coincide. In particular we can find the total age of the universe from (\ref{tc1}): with $(\alpha,\beta)=(0.45,1.4\times 10^{-4})$ calculate
\[\int_0^\infty\frac{dS}{(\beta+\alpha S+S^4)^{1/2}}=3.61,\]
so $\tau_F=3.61/(a_0H)$. At the present time $S=1$ and correspondingly
\[\int_0^1\frac{dS}{(\beta+\alpha S+S^4)^{1/2}}=2.66,\]
so that now $\tau=\tau_0=2.66/(a_0H)$ and the universe has used up $74\%$ of its total conformal time -- just $26\%$ is left!

In terms of proper time, (\ref{tc4}) will give the present age of the universe: integrate to $S=1$ to obtain $t=1.38\times 10^{10}$ years, as expected.

Equations (\ref{tc3}, \ref{tc4}) are exact but there are two simple approximations giving an answer in elementary functions:
\begin{itemize}
 \item at late times we can neglect the radiation as compared to the dust and (\ref{tc4}) is solved by
 \be\label{tc5} S=\alpha^{1/3}(\sinh(3Ht/2))^{2/3},\ee
 so that as expected $S\sim e^{Ht}$ at large $t$. (This solution is actually on Wikipedia, with the claim that it is good for times $t>10^7$ years.)
 
 It's worth doing a little more with this expression. Quite soon in the history of the universe, the exponential expansion will swamp the density term:
 \[\frac{\kappa\rho_m}{H^2}=\frac{\alpha a_0^3}{a^3}<10^{-2}\mbox{  when  }a>(100\alpha)^{1/3}a_0\mbox{  i.e. }a>3.6a_0,\]
 and then we can expand $S$ in (\ref{tc5}) and integrate for $\tau$:
 \[\int_t^\infty a^{-1}dt=a_0^{-1}\int_t^\infty S^{-1}dt=a_0^{-1}\int_t^\infty \alpha^{-1/3}e^{-Ht}dt=a_0^{-1}\alpha^{-1/3}H^{-1}e^{-Ht}\]\[=\int_\tau^{\tau_F}d\tau=(\tau_F-\tau),\]
 so that late on, with $a>4a_0$ or so, we have the approximation
 \be\label{tc7} e^{-Ht}=\alpha^{1/3}a_0H(\tau_F-\tau).\ee
 This tells us that everything from about $t>10t_0$ is bunched up very close to $\scri^+$ in conformal time. We saw above that $(\tau_F-\tau_0)/\tau_F$, the fraction of conformal time ahead of us now, is about $0.26$. If we 
 replace $\tau_0$, which corresponds to $t_0$, by $\tau_1$ corresponding to $10t_0$ this fraction drops to about $10^{-4}$.

 As a concrete instance of this, the Higgs mechanism is supposed to kick in between $10^{-12}$ and $10^{-6}$ seconds after the Bang; from (\ref{tc6}) below that is $a_0H\tau$ in the range $10^{-13}-10^{-10}$. The same conformal time \emph{before} $\tau_F$ 
 corresponds to $t=4-5\times 10^{11}$ years, which is about thirty times the current age of the universe.

 \item at early times we can neglect the cosmological term as compared to the densities and follow \cite{bau}. There is an early time when the two densities are equal:
 \[\rho_m=\rho_\gamma=\rho_{eq}\mbox{ say, when   }1=\frac{B}{Aa}=\frac{\beta}{\alpha}.\frac{a_0}{a},\]
 so with $\beta/\alpha\sim 3\times 10^{-4}$ this is when $a/a_0\sim 3\times 10^{-4}$ (redshift about 3000). Suppose this happens when $t,\tau$ and $a$ are respectively $t_{eq},\tau_{eq}$ and $a_{eq}$. From the Friedmann equation, 
 omitting $H$ and using conformal time we calculate
 \[\frac{d^2a}{d\tau^2}=a^2\ddot{a}+a\dot{a}^2=\frac{\kappa A}{2},\]
 which is constant, so that this integrates to
 \[a=\frac{\kappa A}{4}\tau^2+C\tau,\]
 for a constant of integration $C$ which can be fixed by the Friedmann equation
 \[\left(\frac{da}{d\tau}\right)^2=a^2\dot{a}^2=\kappa(Aa+B),\]
 whence $C^2=\kappa B$.
 
 At equality we have
 \[a_{eq}=\frac{B}{A},\;\; \rho_{eq}=\frac{A}{a_{eq}^3}=\frac{A^4}{B^3}\]
 so that the expression for $a(\tau)$ can be rewritten
 \be\label{tc6}
 a(\tau)=a_{eq}\left(\left(\frac{\tau}{\tau_*}\right)^2+2\left(\frac{\tau}{\tau_*}\right)\right),\ee
 with $\tau_*=(\kappa \rho_{eq}a_{eq}^2/4)^{-1/2}$. Note that $\tau_{eq}$ is defined by
 \[a(\tau_{eq})=a_{eq}\mbox{  so  }\left(\frac{\tau_{eq}}{\tau_*}\right)^2+2\left(\frac{\tau_{eq}}{\tau_*}\right)=1\]
 and $\tau_{eq}=\tau_*(\sqrt{2}-1)$. It is a simple matter to relate $t$ to $\tau$ in this case, since
 \[dt=ad\tau\mbox{  so that  }t=\tau_*a_{eq}\left(\frac13\left(\frac{\tau}{\tau_*}\right)^3+\left(\frac{\tau}{\tau_*}\right)^2\right).\]
 \item Finally it's worth noting that the ranges of applicability of these two approximations overlap. Neglecting $\rho_\gamma/\rho_m$ to obtain the first is equivalent to neglecting $(\beta a_0)/(\alpha a)$ 
 and this is less than $10^{-2}$ for 
 $a>3\times 10^{-2}a_0$. Neglecting $H^2/(\kappa\rho_m)$ to obtain the second is equivalent to neglecting $a^3/(\alpha a_0^3)$ and this is less than $10^{-2}$ for $a<10^{-1}a_0$.
 
\end{itemize}

\subsubsection{Conformally rescaling FLRW models}\label{SSS2.1.2}

It is convenient to conformally-rescale an FLRW metric with $\Omega=a^{-1}$. Then
\[\hg=a^{-2}g=d\tau^2-d\sigma^2_k,\]
where $d\tau=dt/a$, which we continue to take as the definition of conformal time. This metric is evidently everywhere regular. It's conformally-flat (since FLRW is known to be) and is conformal to 
part of the Einstein static cylinder (this is clear for $k=1$ and not hard to see for $k=0$ or $-1$). 

Provided $a(t)^{-1}$ is integrable at $t=0$, we may choose the origins of $\tau$ and $t$ to coincide. By (\ref{c9}) this holds for $\gamma>2/3$ which includes 
the usual physically allowed range. If one allows $\gamma\leq 2/3$ then one has initial singularities which, in the language of Section 1.2.7 are `conformally-infinitely far away'. 

With positive $\lambda$, $a(t)^{-1}$ is integrable towards infinity by (\ref{c4}) (evidently this holds for other matter models too). Thus there will always be a future boundary $\scri^+$ and assuming the origins of $t$ 
and $\tau$ coincide, it will be located at a final value
\[\tau=\tau_F=\int_0^\infty\frac{dt}{a(t)}.\]
These models have a finite life-time in conformal time.

\medskip 

Metrics like this are widely referred to as `asymptotically de Sitter' even though the spatial metric, which becomes the metric of $\scri^+$, contains $k$ which can have any value (while for de Sitter $k=1$).

\subsubsection{Rescaling at the Bang}\label{SSS2.1.3}
The process of conformal rescaling, provided the initial singularity was not conformally-infinitely remote, extended the space-time to a smooth surface $t=0=\tau$ at which the extended metric was smooth. This is despite 
the fact that the scale factor $a(t)\sim t^{2/3\gamma}$ is never differentiable for the physical range of $\gamma$ as a function of $t$ at $t=0$. In terms of $\tau$, $a\sim\tau^{2/(3\gamma-2)}$. This is 
differentiable with zero derivative for $\gamma<4/3$, differentiable with nonzero derivative for $\gamma=4/3$, the radiation value, and not differentiable for $\gamma>4/3$. When we seek 
to extend these calculations to solutions with non-zero Weyl tensor, as we shall, we need to remember this.

\subsection{Conformal Einstein equations}\label{SS2.2}
The example of FLRW extends at both ends. Motivated by this, we might seek to pose an initial value problem for a cosmological model either with data at $\scri^+$ or with data at an initial singularity. This would need 
an extension of the EFEs as these surfaces are not in space-time but are added as boundaries. This extension will be the \emph{Conformal Einstein equations}, which are the Einstein equations for $g$ written as equations for $\hg$ and either 
$\Omega$ or $\Omega^{-1}$, whichever extends to the data surface under consideration.
\subsubsection{Data at future infinity}\label{SSS2.2.1}
With a positive cosmological constant one expects, and in many cases can prove, that there is automatically a conformal rescaling that adds $\scri^+$ as a future boundary. It is therefore natural to contemplate posing 
an IVP with data at $\scri^+$ and this was studied thirty years ago by Friedrich, \cite{f1} for the vacuum equations (with $\lambda>0$) and \cite{f2} for the Einstein-Yang-Mills equations with $\lambda>0$ (see also the 
monograph \cite{kro}). 
The method is to find a system of conformal Einstein equations which are regular at $\scri^+$ and equivalent to a symmetric hyperbolic system, for which existence and uniqueness theorems are available. 

In the simpler case 
of vacuum, the data at $\scri^+$ consist of two symmetric tensors $(h_{ij}, E_{ij})$ where $h_{ij}$ is a Riemannian metric on $\scri^+$ and $E_{ij}$, which is trace-free and divergence-free in the metric covariant derivative 
determined by $h_{ij}$, is the derivative normal to $\scri^+$ of the electric part of the Weyl tensor at $\scri^+$ (recall that the Weyl tensor itself is necessarily zero at $\scri^+$, but $E_{ij}$ can be thought of as 
the gravitational radiation data at $\scri^+$). There is a `gauge freedom':
\be\label{f1}(h_{ij},E_{ij})\rightarrow(\tilde{h}_{ij},\tilde{E}_{ij})=(\theta^2h_{ij},\theta^{-1}E_{ij}),\ee
for any real, positive function $\theta$ on $\scri^+$ which automatically preserves the conditions on $E$. This corresponds to a freedom to change the conformal factor $\Omega$ which is being used to add 
$\scri^+$ as a boundary.
In cases with matter there will also 
be data for the matter variables at $\scri^+$.

\medskip 

A proposal by Starobinsky \cite{s1}, proved in some cases of interest by Rendall \cite{r1}, was that, in the presence of a cosmological constant $\lambda=3H^2$, the space-time metric should have an expansion of the form
\be\label{st1}g=dT^2-e^{2HT}(a_{ij}+e^{-2HT}b_{ij}+e^{-3HT}c_{ij}+\ldots)dx^idx^j,\ee
where the spatial metrics $a_{ij},b_{ij},\ldots$ are time independent. This kind of expansion resembles the expansion of the Poincar\'e metric in the \emph{ambient metric construction} 
of Riemannian geometers \cite{fg}. The correspondence is not exact: in the ambient metric expansion only even powers of the defining function of 
the boundary appear. If we take $e^{-HT}$ to be the defining function of the boundary, so that $\scri^+$ is at $T=\infty$, then $a_{ij}$ is the metric of $\scri^+$, $b_{ij}$ is linearly related to the Ricci tensor of $\scri^+$:
\be\label{s2}
H^2b_{ij}=-(R^{(a)}_{ij}-\frac14R^{(a)}a_{ij}),\ee
where $R^{(a)}_{ij},R^{(a)}$ are respectively the Ricci tensor and Ricci scalar of the metric $a_{ij}$, and $c_{ij}$ is proportional to the tensor $E_{ij}$ which is free 
data in Friedrich's 
construction. The freedom (\ref{f1}) corresponds to a freedom to shift the origin in the $T$-coordinate:
\[T\rightarrow\tilde{T}= T-\frac{1}{H}\log\theta,\]
which must be accompanied by a change in the comoving coordinates $x^i$.

\subsubsection{Data at the Bang: Bianchi-III models}\label{SSS2.2.2}
As a simple example with data at the initial singularity, we consider radiation fluid solutions with cosmological constant $\lambda$ and metric of the Bianchi-III form. This is one of the class of spatially-homogeneous but 
anisotropic metrics for which the EFEs therefore 
reduce to ODEs. For diagonal Bianchi-III the space-time metric can be parametrised in terms of two functions of time $(a(t),b(t))$ as follows:
\be
g=dt^2-(ab^2)^2dz^2-(ab^{-1})^2(d\theta^2+\sinh^2\theta d\phi^2).
\label{ks1}
\ee
The space-sections have a time-dependent product metric on $\mathbb{R}\times\mathbb{H}^2$. Note that the determinant of the spatial metric is a function of $\theta$ times $a^3$ so $a$ has the character of the FLRW scale-factor, and 
$b$ measures anisotropy of the spatial metric. We introduce conformal time $\tau$ 
via $d\tau=dt/a$ and rescale the metric: 
\[\hg:=a^{-2}g=d\tau^2-b^4dz^2-b^{-2}(d\theta^2+\sinh^2\theta d\phi^2).\]
The conservation equation with a radiation fluid integrates as before to give $\rho=\frac{3m}{8\pi G}a^{-4}$ for some positive real constant $m$. 
The Einstein equations can be written as the system
\begin{eqnarray}
3\frac{b''}{b}+6\frac{a'b'}{ab}-3\left(\frac{b'}{b}\right)^2+b^2&=&0\label{ks2}\\
3\frac{a''}{a}+3\left(\frac{b'}{b}\right)^2-b^2-2\lambda a^2&=&0\label{ks3}
\end{eqnarray}
where prime is $d/d\tau$ and these have been simplified with the aid of the Hamiltonian constraint, which in turn can be written as
\be
3\left(\frac{a'}{a}\right)^2= 3\left(\frac{b'}{b}\right)^2+b^2+\frac{m}{a^2}+\lambda a^2.
\label{ks5}
\ee
With $(m,\lambda)$ fixed, these equations give a well-posed IVP with data $(a,b,a',b')$ subject to (\ref{ks5}), which is preserved by the evolution. However we want to give data \emph{at the Bang}. 
We shall see that these equations also have solutions with initial data $(a,a',b,b')=(0,\sqrt{m/3},b_0,0)$ at (say) $\tau=0$. Since $a=0$ there, this is an initial singularity, and we are working with the conformal 
Einstein equations (since the Einstein equations are not defined at the singularity) but the situation is different from data at $\scri^+$. Also, since we wish to give data at the Bang, we are constrained to give less of it.

From (\ref{ks5}), the cosmology expands forever (assuming it does so initially) so that $a'>0$ always. The metric cannot recollapse, as we see by introducing
\be
Q=3\left(\frac{b'}{b}\right)^2+b^2,
\label{ks8}
\ee
which is manifestly non-negative. Then by (\ref{ks2})
\[Q'=-12\frac{a'}{a}\left(\frac{b'}{b}\right)^2\leq 0\]
so that, with the chosen initial conditions, 
\[0\leq Q\leq b_0^2.\]
Therefore $b$ and $\frac{b'}{b}$ are bounded for all time. Write (\ref{ks5}) as
\be
3\left(\frac{a'}{a}\right)^2=\frac{m}{a^2}+\lambda a^2+Q.
\label{ks6}
\ee
Thus $\frac{a'}{a}$ is bounded as long as $a$ is and solutions exist until $a$ diverges, which will define $\scri^+$. This happens after 
finite conformal time, as we see by comparing $a$ with the solution $L$ of the equation
\[3(L')^2=m+\lambda L^4,\;\;L(0)=0,\;\;L'(0)>0.\]
Then $a\geq L$ but $L$ diverges in a time
\[\sqrt{3}\int_0^\infty\frac{dL}{(m+\lambda L^4)^{1/2}},\]
which is finite, so $a$ goes to infinity at, say, $\tau=\tau_F<\infty$. 

\medskip

We shall return to the asymptotic form after proving existence
of solutions with data as claimed. For this, we put the system of Einstein equations into a first-order Fuchsian form. Set
\be
a=\tau e^U,\;\;b=e^\Sigma,
\label{ks7}
\ee
when (\ref{ks2}) and (\ref{ks3}) become
\begin{eqnarray*}
\Sigma'&=&Z\\
U'&=&W\\
Z'&=&-\frac{2}{\tau}Z-2WZ-\frac{1}{3}\;e^{2\Sigma}\\
W'&=&-\frac{2}{\tau}W-Z^2-W^2+\frac{1}{3}\;e^{2\Sigma}+\frac{2\lambda}{3}\tau^2e^{2U}
\end{eqnarray*}
while the Hamiltonian constraint becomes
\[m=e^{2U}\left(3+6\tau W+\tau^2(3W^2-3Z^2-e^{2\Sigma}-\lambda\tau^2e^{2U})\right),\]
which can be interpreted just as the statement that the right-hand-side is constant, which evaluation at $\tau=0$ shows to be positive. (In this setting, the momentum constraint is vacuously satisfied.)

The system is of the form
\[\frac{d{\bf{X}}}{d\tau}=\frac{1}{\tau}{\bf{M}}{\bf{X}}+{\bf{B}}({\bf{X}},\tau),\]
where ${\bf{X}}(\tau)=(\Sigma,U,Z,W)^T$ is the vector of unknowns, ${\bf{M}}$ is a constant matrix and ${\bf{B}}$ is smooth (in fact analytic) in its arguments.

This system is singular as it has a pole in $\tau$ at the initial singularity. It is a \emph{first-order Fuchsian system} and there is an existence theorem for such systems \cite{RS}: if the matrix ${\bf{M}}$ has no positive integer eigenvalues 
then the problem is well-posed for data annihilated by ${\bf{M}}$. Here ${\bf{M}}$ clearly has no positive eigenvalues and the allowed data take the form $(\Sigma, U,Z,W)=(\Sigma_0,U_0,0,0)$ at $\tau=0$.

Thus the solution exists at least for an interval in $\tau$ and then the discussion above shows that existence continues until $a$ diverges. There is a 2-parameter family of solutions: the datum $U_0$ is equivalently $a'(0)$ 
and tied to $m$ by the Hamiltonian constraint; the datum $\Sigma_0$ determines $b_0=b(0)$ and therefore determines the metric on the initial singularity, which is
\[\hat{h}=-b_0^4dz^2-b_0^{-2}(d\theta^2+\sinh^2\theta d\phi^2).\]

\medskip

As we go towards $\scri^+$ in this model, $a$ diverges and from (\ref{ks6}) we see that
\[\frac{a'}{a^2}=H+O(a^{-2}),\]
taking the positive square root (as is allowed). Thus there is a simple pole in $a$:
\[a(\tau)=\frac{1}{H(\tau_F-\tau)}+O(1),\]
where $\tau\rightarrow\tau_F$ as $t\rightarrow\infty$. Solving for $a$ in proper-time $t$ gives
\[a=e^{Ht}(1+0(e^{-Ht})),\]
and we are obtaining the Starobinski expansion (\ref{st1}). It's an exercise to solve (\ref{ks1})-(\ref{ks2}) for the first few terms in a series in $(\tau_F-\tau)$ and confirm the 
expansion (\ref{st1}). In particular the metric of $\scri^+$ is fixed by the $O(1)$ term in the expansion of $b$, and the $O((\tau_F-\tau)^3)$ term in $b$ is not fixed by the metric of $\scri^+$ but is the free data 
corresponding to Friedrich's $E_{ij}$.

\medskip

There are examples like this for all Class A Bianchi types in \cite{t2}.

\subsubsection{Contrasting the two cases}\label{SSS2.2.3}
For the IVP at $\scri^+$, the conformal Einstein equations of Friedrich \cite{f1} give a regular symmetric hyperbolic system requiring two tensors for the data, namely 
the metric of $\scri^+$ and the data for the gravitational radiation. 
For the IVP at the initial singularity, the conformal Einstein equations give a singular but Fuchsian system and require only one tensor for data, the metric of the Bang. 
Function counting therefore indicates that initial singularities at which one can pose an IVP have less free data and must be special, which is confirmed by the observation that in particular they must have finite Weyl tensor: 
in the rescaled space-time there is no singularity so 
$\hat{C}_{abc}^{\;\;\;\;\;d}$ in particular is regular, but $C_{abc}^{\;\;\;\;\;d}=\hat{C}_{abc}^{\;\;\;\;\;d}$ so this must also be regular, though this may be hard to detect. This will be the topic of a later 
section, but note the consequence that in general the solutions with data given at $\scri^+$ will not evolve back to give initial singularities with finite Weyl tensor.

\section{Conformal Cyclic Cosmology}\label{S3}
Now we can introduce CCC, motivating it as an interaction of an earlier proposal of Penrose, the \emph{Weyl Curvature hypothesis} \cite{p3}, with the discovery that there is a positive cosmological constant in the world.
\subsection{The Weyl Curvature Hypothesis}\label{SS3.1}
In \cite{p3} Penrose gave an argument that the Big Bang, viewed as a singularity of a Lorentzian manifold, was much more special than the Einstein equations alone could explain. Very simply put, 
Penrose's argument is that near the Bang the matter content of the Universe was close to thermal equilibrium, in other words in a state of very high, possibly maximum, entropy; however the Universe as a 
whole could not have been in 
a state of maximum entropy since it is an everyday experience that entropy continues to increase today -- there is a Second Law of Thermodynamics. Thus some other component of the Universe must have been in a state 
of low entropy, equivalently in a special state, and this other component could only be the gravity or equivalently the geometry. Via the Einstein equations, the matter is point-wise tied to the Ricci tensor so, Penrose argues, 
it must be the Weyl tensor that was special at the Bang. 

Penrose gave quantitative force to the argument by a calculation of the entropy of the part of the Universe inside our past light cone and back to the Bang. His estimate 
for this in 1979 was $10^{88}k$, $k$ being Boltzmann's constant, and he later raised the estimate to $10^{111}k$ when the consensus took hold that most galaxies have supermassive black holes at their core which contribute to this sum by their Bekenstein-Hawking entropy. While 
this would seem to be an impressively large number, Penrose observes that if all the matter inside our past light cone was collapsed into black holes the corresponding figure would be $10^{123}k$, which is already vastly 
greater. Following Boltzmann, whose fondness for the formula $S=k\log W$ was such that he had it engraved on his tomb, Penrose observes that the volume $W$ in phase space corresponding to the actual Universe is even more 
vastly smaller than the apparently available volume: this is $10^{10^{111}}$ as a fraction of $10^{10^{123}}$ and is possibly the smallest number ever contemplated in physics\footnote{Although in 
\cite{is} one finds an estimate of the chance of obtaining the observed universe from inflation 
as one in $10^{10^{56}}$, which is about the same order of order of order of magnitude.}, but this is the fraction of phase space that the Creator's pin 
had to hit to get the universe we have, a fact which needs explanation.

\medskip

There is no agreed measure of gravitational entropy but in \cite{p3} Penrose argued for a connection between it and the Weyl tensor, and recalled how this worked for linear spin-2 theory in Minkowski space. After \cite{p3} 
appeared, various suggestions were made for definitions of gravitational entropy in terms of scalar invariants of curvature but these have the wrong differential order (in terms of the number of derivatives of the metric 
arising in them) to correspond to linear theory where one knows the answer. A more recent discussion of these definitions is given in \cite{pl}. An attempt to follow Penrose's suggestion more closely for cosmologies close to FRW 
was made in \cite{tm}. That definition has at least one good property but it isn't clear how to extend the definition to cosmologies further away from FRW.

\medskip 

Even without a universal definition of gravitational entropy, Penrose took the view in \cite{p3} that the connection with the Weyl tensor was clear and proposed that the simplest conjecture to make was that 

\medskip

\emph{the Weyl tensor 
$C_{abcd}$ is zero at any initial singularity}. 

\medskip

\noindent This is {\bf{the Weyl Curvature Hypothesis}}. The force of `initial' in the formulation is that this property could not hold for singularities formed in collapse to black holes since these are likely to have the character of the Schwarzschild 
singularity at which the Weyl tensor is certainly singular (since the Riemann tensor is singular but the Ricci tensor is zero). If the Universe were to recollapse to a Big Crunch singularity then this would be at least as bad as 
the Schwarzschild singularity and could be more like the chaotic Mixmaster singularities, at that time conjectured (and now known \cite{hr}) to occur in vacuum Bianchi-IX collapse. Penrose in \cite{p3} wrote the 
Weyl tensor as here with all indices down, i.e. as $C_{abcd}$, which seemed to suggest that he would be content to have $C_{abc}^{\;\;\;\;\;d}$ finite (when lowering the index would lead it to vanish where the metric vanished). It is 
possible to make this interpretation but at the time he told me that that was a finer distinction than he wanted to make.

\medskip

Whether the Weyl Curvature Hypothesis is that $C_{abc}^{\;\;\;\;\;d}$ vanishes or is finite at the initial singularity is a smaller question than that of how one is to tell: at the Bang, the Riemann tensor is singular as is the 
metric and therefore the metric connection, so how is one to tell that the Weyl tensor, whose components are ten of the twenty components of the Riemann tensor, is finite? Scalar invariants of curvature bring in the metric or 
the volume-form so can mislead; likewise, components in a parallelly-propagated frame involve the metric connection.

Another question that Penrose considered in \cite{p3} is how do you \emph{cause} something, like vanishing of the Weyl tensor, which happens \emph{at the beginning}? We'll see a nice answer to this below, but at the time he 
speculated that the `correct' theory of quantum gravity, once it was found, might not be invariant under time-reversal.

\subsection{Conformal gauge singularities}\label{SS3.2}

There is a simple way to obtain singular space-times with a singularity at which the Weyl tensor is finite: start with a smooth space-time say $M$ with metric $g_{ab}$ 
and choose a smooth hypersurface $\Sigma$, most commonly space-like, and a function $\cOmega$ which vanishes at $\Sigma$; now rescale the metric -- $g_{ab}\rightarrow \cg_{ab}=\cOmega^2g_{ab}$. The rescaled space-time $\cM$ say is certainly singular at 
$\Sigma$ but its Weyl tensor by (\ref{c1}) is not: $\cC_{abc}^{\;\;\;\;\;d}=C_{abc}^{\;\;\;\;\;d}$ which is smooth. Singularities formed like this have had different names over the years: \emph{isotropic} in \cite{gw}, 
since a co-moving volume shrinks at the same rate in all directions approaching the singularity, and \emph{conformally compactifiable}, by analogy with behaviour of WAS space-times at $\scri$, but I think the best name 
is \emph{conformal gauge singularities} (used in \cite{tl} but due to Christian L\"ubbe) by analogy with a coordinate singularity, since a change of conformal gauge removes the singularity.

\medskip

For the applications that we have in mind, we require $\Sigma$ to be space-like in $M$ but the example of FRW (Section 2.1.2) reminds us that we may not want to assume that $\cOmega$ is smooth at $\Sigma$. Now two questions arise:
\begin{enumerate}
 \item Is there a well-posed initial value problem with data at $\Sigma$ for the Einstein equations in $\cM$? This will depend on the matter model considered and raises a separate question: what happens if the Weyl tensor is 
 zero rather than just finite?
 \item How general is this class of singularities with finite Weyl tensor? Are there other classes of finite Weyl tensor singularities?
\end{enumerate}
\subsubsection{The IVP for perfect fluids}\label{SSS3.2.1}
Following the calculation in Section 2.2.2, we shouldn't be surprised if there is an IVP but we should expect the evolution equations to have a singularity in the time. I'll do the radiation case as the simplest example 
so the physical energy-momentum tensor is 
\[\cT_{ab}=\frac13\crho(4\cu_a\cu_b-\cg_{ab}),\]
where $\crho$ is the fluid density, diverging at the Bang, and $\cu^a$ is the physical fluid velocity, unit w.r.t. $\cg$. Here I'm introducing a convention that we'll adopt in this chapter: the physical metric after the Bang 
and associated quantities carry a check (more accurately a h\'a$\check{\mathrm{c}}$ek); unhatted and unchecked metrics and tensors will be unphysical. 

It was shown in \cite{gw} that for large classes of perfect fluids with an isotropic singularity that the rescaled fluid flow must be orthogonal to the singularity surface and must therefore be twist-free. Thus the flow 
defines a cosmic time $\tau$, with the freedom to replace it by a function of itself, and we may also introduce comoving space-coordinates. 

To have a conformal factor vanishing at the Bang we take the unphysical metric to be $g_{ab}=\cOmega^{-2}\cg_{ab}$ and then the unphysical fluid velocity is $u^a=\cOmega\cu^a$. Guided by the FRW case we assume that $\cOmega$ 
is smooth at the Bang in the unphysical time-coordinate $\tau$, and we'll actually take $\cOmega=\tau$ which ultimately is justified by deriving a solvable IVP.

The surface $\Sigma$ is at $\tau=0$. It has unit normal $N_a=V^{-1}\tau_{,a}=u_a$ for a positive scalar $V$ and the unphysical metric in comoving coordinates is
\[g=V^{-2}d\tau^2-h_{ij}dx^idx^j.\]
The second fundamental form $K_{ab}$ of the constant $\tau$ foliation and the unphysical acceleration $A_a$ are fixed by
\[\nabla_aN_b=N_aA_b+K_{ab},\]
with
\[N^aA_a=0=N^aK_{ab}.\]
Since also $N_a=V^{-1}\tau_{,a}$ we have
\[\nabla_a\tau_b=V(N_aA_b+K_{ab})+V_aN_b,\]
which must be symmetric so that
\[V_a=V(A_a+N_aV_\tau).\]
Recall (\ref{c2}):
\[\check{R}_{ab}=R_{ab}+2\nabla_a\Upsilon_b+g_{ab}\nabla_c\Upsilon^c-2\Upsilon_a\Upsilon_b+2g_{ab}\Upsilon_c\Upsilon^c,\]
then substituting for $\Upsilon$ and using the physical Einstein equations gives
\be\label{3.1}-\frac{8\pi\crho G\tau^2}{c^2}(4u_au_b-g_{ab})=R_{ab}-\frac{V^2}{\tau^2}(4N_aN_b-g_{ab})+\frac{V}{\tau}(K_{ab}-Kg_{ab}+2N_{(a}A_{b)}+N_aN_bV_\tau).\ee
Following \cite{t3}, from the conservation equation in physical space-time and the freedom to redefine $\tau$ we obtain a simple expression for the density: for a radiation fluid we may suppose
\[\frac{8\pi\crho G}{c^2}=\frac{V^4}{\tau^4}.\]
The conservation equation also gives
\be\label{e7}\frac{\partial V}{\partial\tau}=-\frac13K.\ee
With the given expression for the density, (\ref{3.1}) becomes
\be\label{3.2}R_{ab}-\frac{V^2(1-V^2)}{\tau^2}(4N_aN_b-g_{ab})+\frac{V}{\tau}(K_{ab}-Kg_{ab}+2N_{(a}A_{b)}+N_aN_bV_\tau)=0.\ee
We decompose (\ref{3.2}) in the standard $(3+1)$ manner to obtain an evolution equation and two constraints. The evolution equation is
\be\label{e8}
\mathcal{L}_NK_{ab}=R^{(h)}_{ab}-KK_{ab}+2K_{ac}K_b^{\;\;c}+\nabla^{(h)}_aA_b-A_aA_b-2\frac{V}{\tau}K_{ab}-h_{ab}\left(\frac{2VK}{3\tau}+\frac{V^2(1-V^2)}{\tau^2}\right),\ee
where $\nabla^{(h)}_a$ is the metric connection of the space metric $h_{ab}$ and $R^{(h)}_{ab}$ is its Ricci tensor. The constraints are
\be\label{e9}G_{00}=-\frac12(R^{(h)}-K^2+K_{ab}K^{ab})=-2\frac{VK}{\tau}-3\frac{V^2(1-V^2)}{\tau^2}\ee
and
\be\label{e10}
G_{0a}=\nabla^{(h)}_bK_a^{\;\;b}-\nabla^{(h)}_aK=\frac{2}{\tau}\nabla^{(h)}_aV.\ee
It is straightforward to check that the constraints are preserved by the evolution. The unknowns are $(V,h_{ab},K_{ab})$ with evolution determined by (\ref{e7}), (\ref{e8}) and the definition
\[\mathcal{L}_Nh_{ab}=2K_{ab}.\]
The system isn't yet in Fuchsian form, since (\ref{e8}) has second-order poles in $\tau$ but this problem can be solved by some redefinitions of variables. One can read off constraints on data at $\tau=0$: from (\ref{e10}), $V$ 
must be constant; from (\ref{e8}) $K_{ab}$ must vanish initially and the constant value of $V$ must be one; nothing new comes from (\ref{e9}). The details are in \cite{at1} (see also \cite{n1}, \cite{cn}) and the conclusion is

\medskip

\emph{the solution exists, is unique and depends continuously on the data which are just the 3-metric $h^{(0)}_{ab}$ of the initial surface.}

\medskip

One readily obtains the first few terms in power series in $\tau$:
\[h_{ab}=h^{(0)}_{ab}+\tau^2k_{ab}+\mbox{h.o.},\;\;K_{ab}=\tau k_{ab}+\mbox{h.o.},\;\;V=1-\frac{k}{6}\tau^2+\mbox{h.o.},\]
with
\[k_{ab}=\frac13(R^{(0)}_{ab}-\frac16R^{(0)}h^{(0)}_{ab}),\;\;k=\frac16R^{(0)}.\]
Inductively it's clear that $h_{ab}$ and $V$, and therefore the space-time metric, are series in even powers of $\tau$. Inhomogeneities in $V$ and therefore in the density arise if the initial 3-Ricci scalar is nonconstant. 
Since the initial $K_{ab}$ must be zero, it follows that the magnetic part of the initial Weyl tensor must 
be zero. It was observed in \cite{gw} that the electric part of the initial Weyl tensor is proportional to the trace-free part of the initial 3-Ricci tensor. Thus if the whole of the initial Weyl tensor is zero then 
the initial 3-metric has vanishing trace-free Ricci tensor, so is Einstein and so is data for FLRW. Uniqueness of solution implies that the solution is then FLRW and in this case

\medskip

\emph{if the Weyl tensor is zero initially then it is always zero.}

\medskip

In the radiation case this was considered in \cite{new1}, and, along with other polytopic fluids, was shown in \cite{at1} still with this Weyl tensor property: if it vanishes initially then it is always zero. This is rather a strong 
property so it is natural to consider other matter models.
\subsubsection{The IVP for massless Einstein-Vlasov}\label{SSS3.2.2}
This case has trace-free physical $\check{T}_{ab}$ so we'll again take $\cOmega=\tau$ and (\ref{3.2}) is replaced by
\be\label{3.7}R_{ab}-\frac{V^2}{\tau^2}(4N_aN_b-g_{ab})+\frac{V}{\tau}(K_{ab}-Kg_{ab}+2N_{(a}A_{b)}+N_aN_bV_\tau)+\frac{\kappa V}{\tau^2\sqrt{h}}\int\frac{fp_ap_b}{p_0}d^3p=0.\ee
Here $\kappa=8\pi G/c^2$, $p_0=(h^{ij}p_ip_j)^{1/2}$ and $h=\mbox{det}h_{ij}$. One also has the Vlasov equation (\ref{v2})
\be\label{3.8}
V^2\frac{\partial f}{\partial\tau}-h^{ij}p_i\frac{\partial f}{\partial x^j}-\frac12\left((p_0)^2\partial_iV^2-p_mp_n\partial_ih^{mn}\right)\frac{\partial f}{\partial p_j}=0.\ee
The Vlasov equation has no singularity at $\tau=0$ but the rescaled Einstein equation (\ref{3.7}) does and one needs to split it as before into constraints and evolution and obtain the first-order Fuchsian form. This can 
be done (\cite{a}, \cite{at2}) but one may obtain the Fuchsian conditions on the data by a simple-minded approach of seeking power-series solutions.

First one can adjust the conformal gauge to set $V=1$ at $\tau=0$; then (\ref{3.7}) at $O(\tau^{-2})$ gives conditions on the initial distribution function $f^0(x^i,p_j)$ and initial metric $h^0_{ij}$:
\[\int p_if^0(x^i,p_j)d^3p=0,\;\;h^0_{ij}=\frac{1}{3\sqrt{h^0}}\int\frac{p_ip_jf^0(x^k,p_k)}{(h^{0mn}p_mp_n)^{1/2}}d^3p.\]
The first of these is a `vanishing dipole' condition on the initial distribution function which has the effect of making the time-like eigenvector of the stress-tensor, which can be thought of as a mean matter velocity vector, 
orthogonal to the singularity surface. The second has the appearance of a constraint relating $f^0$ and $h^0_{ij}$ but in fact it's an equation, determining $h^0_{ij}$ given suitable $f^0$. This can be seen by consideration of the problem:

\medskip

\emph{find $h^{ij}$ which minimises $F(h^{ij}):=(\mbox{det}(h^{mn}))^{-1/6}\int f(x^k,p_k)(h^{ij}p_ip_j)^{1/2}d^3p$.}

\medskip

It's straightforward to show that this problem has a unique minimum given non-negative $f$, compactly supported in $p$, and the Euler-Lagrange equations for the minimum are the desired constraint. From the minimisation 
description, this initial metric can be seen to be smooth if $f^0$ is.

One has more Fuchsian constraints from (\ref{3.7}) at $O(\tau^{-1})$ and these determine the initial value of the second fundamental form, $K^0_{ij}$ from the third and fourth moments of $f^0$. Essentially one 
obtains an equation of the form
\[M_{ij}^{\;\;\;\;mn}K^0_{mn}=N_{ij}\]
where $M_{ijmn},N_{ij}$ are obtained from
\[\chi_{ijmn}=\int \frac{p_ip_jp_mp_nf}{(h^{rs}p_rp_s)^{3/2}}d^3p\mbox{  and  }\chi_{ijk}=\int \frac{p_ip_jp_kf}{(h^{rs}p_rp_s)}d^3p,\]
(see \cite{a,at2} for more detail).

The data is strikingly different from the perfect fluid case -- there one gave the initial metric with no extra data for the fluid while here one gives the distribution function and this fixes the geometric data. 
It is also the case that now the Weyl tensor can be zero initially but become nonzero later. In a sense it emerges from the higher multipoles of $f^0$.

\subsubsection{Other matter models}\label{SSS3.2.3}
Analysis leading to an equation corresponding to (\ref{3.2}) or (\ref{3.7}) has been done for a range of other matter models \cite{t4}. Thus these cases have been taken far enough to obtain Fuchsian conditions without 
reducing them completely to the form where well-posedness can be proved. These cases include Einstein-scalar-field with potentials, Einstein-Yang-Mills-Vlasov and Einstein-Boltzmann. The last case is interesting as it must 
bridge between Einstein-perfect-fluid and Einstein-Vlasov depending on the behaviour of the scattering cross-section. Cases with massive Vlasov and nonzero scalar curvature have been studied with spatially-homogeneous 
metrics, \cite{t5}.

\subsubsection{Conformal extension through a singularity}\label{SSS3.2.4}
A question which needs to be addressed is how far one can deduce that a finite Weyl tensor singularity is necessarily a conformal gauge singularity. Equivalently what conditions on the Weyl tensor permit a conformal 
extension? This was considered in \cite {tl} (see also \cite{l2},\cite{tl2}). The idea is to suppose that one has a time-like conformal geodesic $\gamma$ which is incomplete because it runs into a singularity BUT along 
$\gamma$ the Weyl curvature and as many derivatives of it as are necessary are bounded in a suitable frame, suitably propagated. In \cite{tl} this was phrased in terms of the calculus of tractors, 
which won't appear in these lectures.
To do it without tractors, recall subsection 1.2.6 where a conformally-invariant way to decompose the Weyl tensor and its derivatives into components was found. With the aid of these we can give a definition for 
boundedness of the Weyl tensor and its derivatives up to any order $k$. This is the answer to the question raised in section 3.1: how do you identify finite Weyl tensor when the Riemann tensor is singular?

\medskip

Now one can translate the result in \cite{tl} to a form without tractors:

\medskip

\noindent\emph{ Let $\gamma:[0,\tau_F) \to M$ be the final segment of an incomplete time-like conformal geodesic in $(M,g)$, such that $b_a$ is bounded in $[0,\tau_F) $. Let $W \subset M$ be a neighbourhood of $\gamma[0,\tau_F) $ in which the 
strong causality condition holds. Let $\{e_\beta \}$ be a Weyl propagated orthonormal frame along $\gamma$.} 

\noindent \emph{(i) If the Weyl tensor and its derivative have uniformly bounded norms along $\gamma$ with respect to $\{e_\beta \}$ then there  exists a neighbourhood $U$ of 
$\gamma[0,\tau_F)$ with $\overline{U} \subset W $ and a diffeomorphism $\psi: V \subset \mathbb{R}^4 \to U $. }

\noindent\emph{ (ii) Suppose that in $U$ the norms of the derivatives of the Weyl curvature up to order $k$ are uniformly bounded then 
there exists a conformally related metric $g$ such that there exists $U^*\supset \overline{U}$ with a $C^k $-extension of $(U, g_{ij}|_U) $ into 
$(U^*, g_{ij}\|_{U^*}) $. }

\noindent\emph{ (iii) The Riemann curvature of $g$ is $C^{k-1} $. }

\noindent \emph{Thus the conformal structure $(M,\cg) $ is locally extendible. }

\subsection{The `outrageous suggestion'}\label{SS3.3}
Since about 1998 the belief that there is a positive cosmological constant has received ever-strengthening support, including that of the Royal Swedish Academy of Sciences. It was apparent early on to any relativist with a knowledge of the theory in subsection 1.2.4 that this 
indicated that the existence of a space-like $\scri^+$ was therefore very likely. One would still need to show that solutions of the Einstein equations with positive $\Lambda$ generically have a smooth $\scri^+$, 
something that was widely believed under the rubric of the `cosmic no-hair conjecture' \cite{gh} and proved in certain cases.

If there does exist a space-like $\scri^+$ then under weak conditions the Weyl tensor will vanish there and the conformal metric will extend through. On the other hand, ever since 1979 Penrose had wanted a mechanism to cause 
the universe to have an initial singularity at which the Weyl tensor was zero. Now one presents itself, and this led Penrose in about 2005 to his `outrageous suggestion' which is CCC: that the initial singularity should be a conformal gauge 
singularity so 
that the conformal metric could be extended through it and, conformally, the Big Bang of one cosmology could be identified with the $\scri^+$ of an earlier one. He called the different phases `aeons' and there would be a 
well-defined conformal metric extending from aeon to aeon, while each aeon would have a physical metric related to the overall conformal metric by a conformal factor that cycled from zero to infinity. The 
picture is not intended to be \emph{periodic} but it does go through cycles.

\medskip

A very simple example is provided by the FLRW metric with source a radiation fluid. We dealt with this in section \ref{SS2.1}: the metric is
\[g=dt^2-(a(t))^2d\sigma_k^2,\]
where $d\sigma_k^2$ is one of the three 3-dimensional Einstein metrics. The conservation equation is integrated by $\rho a^4=4\pi \mu/G$ where $\mu$ is a constant of integration and the Einstein equations reduce to the Friedmann equation 
which can be written in terms of conformal time $\tau$ as in (\ref{c10}):
\be\label{c11}\left(\frac{da}{d\tau}\right)^2=\mu-ka^2+\frac13\lambda a^4.\ee
The RHS of (\ref{c11}) is never zero for $k$ negative or zero, and for $k>0$ it won't be zero provided $\lambda\mu>3/4$. Suppose the RHS is never zero, than $a$ expands from a simple zero at say $\tau=0$ and will diverge at some final 
value $\tau_F$, where in fact it will have a simple pole: $a\sim H(\tau_F-\tau)^{-1}+O(1)$, where $\lambda=3H^2,H>0$.

The conformal metric 
\[\tilde{g}=d\tau^2-d\sigma_k^2\]
is independent of $\tau$ so is oblivious to the passage through $\tau=\tau_F$ but (\ref{c11}) has an attractive symmetry: it is invariant if we replace $a$ by $\tilde{a}=c/a$ for some constant $c$, 
and then $\tilde\lambda=3\mu/c^2$ and $\tilde\mu=\lambda c^2/3$. If 
we take the view that $\lambda$ is to be the same at each cycle then we should take $c^2=3\mu/\lambda$ and then $\mu$ is also the same at each cycle. This gives the simplest model of a CCC: the scale factor runs from 
zero to infinity in a finite amount of conformal time, and when it reaches infinity it is replaced by a constant times its reciprocal. In fact this model is periodic but 
its Weyl tensor is identically zero (it vanishes initially and the source is a radiation fluid so it is always zero by section 3.2.1.). We'll be guided by this model in what follows. Note that the model requires the matter 
to be radiation fluid in the previous aeon: any admixture of dust will spoil the symmetry of (\ref{c11}).

\subsection{The equations of CCC}\label{SS3.4}
By and large we follow \cite{p1} in this subsection. In CCC then there are two space-times representing successive aeons, $\hat{M}$ and $\check{M}$, respectively to the past and the future of a common boundary $\Sigma$, and two space-time metrics, $\hat{g}_{ab}$ for $\hat{M}$ 
and $\check{g}_{ab}$ 
for $\check{M}$. There is a third metric $g_{ab}$ on $M=\hat{M}\cup\check{M}\cup\Sigma$ for which $\Sigma$ is smooth and space-like, and the metrics are conformally related by conformal factors $\hat{\Omega}$ and 
$\check{\Omega}$ according to
\[\hat{g}_{ab}=\hat{\Omega}^2g_{ab}\mbox{  on  }\hat{M};\;\;\check{g}_{ab}=\check{\Omega}^2g_{ab}\mbox{  on  }\check{M}.\]
Furthermore $\hat{\Omega}^{-1}$ and $\check{\Omega}$ are smooth and tend to zero at $\Sigma$. We can call $\Sigma$ the \emph{cross-over surface} and $g_{ab}$ the \emph{cross-over metric}.

At this point we are free to rescale the cross-over metric, say $g_{ab}\rightarrow\Theta^2g_{ab}$ when also $\hat{\Omega}\rightarrow\Theta^{-1}\hat{\Omega}$ and $\check{\Omega}\rightarrow\Theta^{-1}\check{\Omega}$. This changes 
the product $\hat{\Omega}\check{\Omega}\rightarrow\Theta^{-2}\hat{\Omega}\check{\Omega}$ so we refine the definition of the cross-over metric by imposing the condition
\be\label{cc1}\hat{\Omega}\check{\Omega}=-1,\ee
(we have $-1$ here rather than $+1$ since each of $\hat{\Omega}^{-1}$ and $\check{\Omega}$ goes through zero at $\Sigma$). 
This is Penrose's \emph{reciprocal hypothesis} imposed as a gauge condition. As a consequence
\[\check{g}_{ab}=\hat{\Omega}^{-4}\hat{g}_{ab},\]
so that the metric of the later aeon is determined by metric of the earlier aeon and the conformal factor $\hat{\Omega}$. In particular therefore the Einstein tensor and so the energy-momentum tensor in the later aeon 
are determined by the metric of the earlier aeon and $\hat{\Omega}$. Consequently it becomes crucial to find a prescription for a unique preferred $\hat{\Omega}$.

This is a convenient place to introduce a piece of terminology from \cite{p2}: Penrose introduces the one-form
\be\label{cc7} \Pi_a=\frac{\nabla_a\hat{\Omega}}{\hat{\Omega}^2-1}=\frac{\nabla_a\check{\Omega}}{1-\check{\Omega}^2}.\ee
This is well-defined at the cross-over surface, as can be seen from the second expression, and $\Pi_a$ is proportional to $N_a=\nabla_a\check{\Omega}$, the normal to the cross-over surface. We'll need $\Pi_a$ below.

It's part of Penrose's view of CCC that, late in any aeon, the energy-momentum tensor of the matter should become trace-free. This isn't necessary for the appearance of a boundary $\scri^+$ at which the Weyl curvature vanishes -- 
one can for example find dust cosmologies with that property -- but for simplicity we'll make the same assumption, although new physics may be needed to enforce it (see below, section 3.5). Thus the matter late in the previous aeon can be 
combinations of radiation fluid, Maxwell fields, massless Vlasov or some other massless field.

With this assumption, Penrose suggested that one should restrict $\hat\Omega$ by requiring that the cross-over metric $g_{ab}$ have the same scalar curvature as $\hat{g}_{ab}$ near to $\scri^+$ and this is just 
$\hat{R}=4\hat\lambda$, with $\hat{\lambda}$ the cosmological constant in the earlier aeon\footnote{I don't want to make this assumption.}. We adapt (\ref{c3}) to this case: it can be directly written as 
\be\label{cc3}\Box\hat{\Omega}+\frac16R\hat{\Omega}=\frac{2}{3}\hat{\lambda}\hat{\Omega}^3\ee
when it is an equation applying the cross-over d'Alembertian to a function which blows up at $\Sigma$ and $R$ is the scalar curvature of the cross-over metric, as yet unfixed; or by introducing $\phi=\hat{\Omega}^{-1}$, 
so that $\phi$ is smooth through $\scri^+$ with a simple zero, as
\be\label{cc2}\hat{\Box}\phi+\frac{2}{3}\hat{\lambda}\phi=\frac16R\phi^3,\ee
$R$ as before. Note (\ref{cc3}) and (\ref{cc2}) have the form of (\ref{d4}), which is the field equation for the conformal scalar field. In this context, and with $R=4\lambda$, 
Penrose \cite{p1} introduces the term \emph{the phantom field equation} for this equation, and \emph{the phantom field} for $\hat{\Omega}$.

Whatever we choose for $R$, we can expand $\phi$ in the manner of the Starobinski expansion (\ref{st1}):
\be\label{cc4}\phi=\Sigma_{n=1}^\infty\phi_n(x^j)e^{-nHT}\ee
where $\hat{\lambda}=3H^2$ and solve (\ref{cc2}) term by term. We find that $\phi_1$ and $\phi_2$, the first two terms in the series (\ref{cc4}), are freely specifiable and subsequent $\phi_n$ are then uniquely determined. In particular 
\be\label{cc55}2H^2\phi_3=\Delta^{(a)}\phi_1-H^2b\phi_1+\frac16R\phi_1^3\ee
where $\Delta^{(a)}$ is the Laplacian of the metric $a_{ij}$ in the Starobinski expansion (\ref{st1}), and $H^2b=\frac14R^{(a)}$ by (\ref{s2}). Thus evaluating $\phi_3$ requires the choice 
of $R$, which we've deferred, to be made.

Note the role of $\phi_1$ in fixing 
the metric of $\scri^+$: from 
(\ref{st1}) and the definition of $\phi$ the metric of $\scri^+$ is $\phi_1^2a_{ij}$. We'll seek to fix $\phi_1$ shortly but comparison of (\ref{c4}) and (\ref{cc4}) shows that in the FLRW case necessarily $\phi_2=0$. 
Since we are taking the FLRW case for guidance, this 
suggests always taking $\phi_2=0$, which is allowed as it's free data. Penrose makes this assumption in \cite{p1}, where he calls it \emph{the delayed rest-mass hypothesis} for a reason we come to next.

\subsubsection{The delayed rest-mass hypothesis}\label{SSS3.4.1}
Given (\ref{cc1}), the Einstein tensor post-bang, say $\check{G}_{ab}$, can be calculated from the Einstein tensor pre-bang, $\hat{G}_{ab}$ and $\phi=\hat{\Omega}^{-1}$. In particular this is true for the Ricci scalar, but 
while the pre-bang Ricci scalar is $\hat{R}=4\lambda$ the post-bang Ricci scalar $\check{R}$ will have an extra term. This, from the trace of the post-bang energy-momentum tensor, will be an indicator of rest-mass appearing 
post-bang. Penrose \cite{p1} writes
\[\check{R}=4\lambda+8\pi G\mu.\]
The relation between the metrics is $\check{g}_{ab}=\phi^4\hat{g}_{ab}$ so that by (\ref{c3})
\[4\lambda+8\pi G\mu=\check{R}=\phi^{-4}(\hat{R}+6\hat{\Box}\phi^2/\phi^2)=\phi^{-4}(-4\lambda+2R\phi^2+12\phi^{-2}|\hat{\nabla}\phi|^2),\]
using (\ref{cc2}). Some manipulation turns this into the expression at the top of p249 in \cite{p1}. From the expansion (\ref{cc4}) we obtain
\[8\pi G\mu=24H^2(\phi_2/\phi_1^3)e^{3HT}+O(e^{2HT}).\]
This is singular at the Bang (where $T$ is infinite), as is to be expected, but we see that the choice $\phi_2=0$ promoted above sets the leading term to zero -- the choice $\phi_2=0$ delays the rate at which rest-mass appears after the Bang, whence 
Penrose's terminology.

\medskip

Coming back to the issue of fixing $\phi_1$, Penrose in \cite{p1} offers several possible choices without settling for any one. These can be taken to be
\begin{itemize}
 \item $N^aN^b\Phi_{ab}=O(\check{\Omega}^3)$,
 \item $N^aN^b\nabla_aN_b=O(\check{\Omega}^2)$,
 \item $\nabla_a\Pi^a=O(\check{\Omega})$, or
 \item $3\Pi^a\Pi_a-\lambda=O(\check{\Omega}^3)$.
\end{itemize}
Here $\Phi_{ab}$ is minus half the trace-free Ricci tensor. 

 In \cite{t6} I suggested a different one, which is to choose $\phi_1$ so that the metric 
of $\scri^+$, which is $\phi_1^2a_{ij}$, has constant scalar curvature. This amounts to solving the Yamabe problem for $\scri^+$
.The solution of this problem is very often unique, 
so that this prescription has at least that virtue, and it is effectively what one does for FLRW, section 3.3, when one assumes that $\phi$ is a function only of $t$.

In the long run, the choice between these possibilities should be physically motivated.

\subsubsection{A result of L\"ubbe}\label{SSS3.4.2}
If there is to be a physically relevant scalar field after the Bang, one might feel that there should also be one in the previous aeon. In \cite{l1}, L\"ubbe described a way to accomplish this. We recall the definition  
\be\label{ol1}D_{ab}[\phi,g_{cd},\alpha]:=4\phi_a\phi_b-g_{ab}g^{cd}\phi_c\phi_d-2\phi\nabla_a\phi_b+2\phi^2L_{ab}+2\alpha\phi^4g_{ab}\ee
from section 1.3.2. This is associated with the field equation
\be\label{ol4}Q(\phi,g_{ab},\alpha):=\Box\phi+\frac16R\phi-4\alpha\phi^3=0,\ee
and there is the conformal invariance: if $\tilde{\phi}=\Omega^{-1}\phi$ then 
\be\label{ol3}D_{ab}[\tilde{\phi},\tilde{g}_{cd},\alpha]=\Omega^{-2}D_{ab}[\phi,g_{cd},\alpha],\;\;Q(\hat{\phi},\hg_{cd},\alpha)=\Omega^{-3}Q(\phi,g_{ab},\alpha),\ee
and one also has
\[g^{ab}D_{ab}=-2\phi Q(\phi),\;\;\nabla^aD_{ab}=4Q\phi_b-2\phi Q_b.\]
Now suppose one has the Einstein equations with source consisting of cosmological constant $\lambda$, some extra massless fields $T^{\mathrm{ex}}_{ab}$ which are separately conserved, 
and the conformal scalar represented by stress-tensor $D_{ab}$:
\be\label{ol2} G_{ab}=-D_{ab}[\phi,g_{cd},\alpha]-\kappa T^{\mathrm{ex}}_{ab}-\lambda g_{ab}.\ee
The trace of this gives $R=4\lambda$ and L\"ubbe remarks that, when $R=4\lambda$ we have by (\ref{ol1})
\[D_{ab}[1,g_{cd},\lambda/4]= -G_{ab}-\lambda g_{ab}. \]
Therefore the EFEs (\ref{ol2}) can be written
\be\label{ol5}D_{ab}[1,g_{cd},\lambda/4]+D_{ab}[\phi,g_{cd},\alpha]-\kappa T^{\mathrm{ex}}_{ab}=0.\ee
Now rescale this:
\[0=\Omega^{-2}\left(D_{ab}[1,g_{cd},\lambda/4]+D_{ab}[\phi,g_{cd},\alpha]-\kappa T^{\mathrm{ex}}_{ab}\right)\]
\[=D_{ab}[\Omega^{-1},\tilde{g}_{cd},\lambda/4]+D_{ab}[\Omega^{-1}\phi,\tilde{g}_{cd},\alpha]-\kappa\widetilde{T}^{\mathrm{ex}}_{ab},\]
where $\widetilde{T}^{\mathrm{ex}}_{ab}=\Omega^{-2}T^{\mathrm{ex}}_{ab}$ which is the correct rescaling for a trace-free stress tensor to preserve conservation.

Now choose $\Omega=\phi$, then this is
\[0=D_{ab}[\phi^{-1},\tilde{g}_{cd},\lambda/4]+D_{ab}[1,\tilde{g}_{cd},\alpha]-\kappa\widetilde{T}^{\mathrm{ex}}_{ab},\]
which we recognise as the EFEs (\ref{ol5}) but for the metric $\hat{g}$ with some other adjustments. L\"ubbe's result is therefore

\medskip

\emph{Given a solution $(\phi,g_{ab},\alpha,\lambda/4,T^{\mathrm{ex}}_{ab})$ of the EFEs (\ref{ol5}), there is another 
with $(\phi^{-1},\tilde{g}=\phi^2g,\lambda/4,\alpha,\widetilde{T}^{\mathrm{ex}}_{ab}=\phi^{-2}T^{\mathrm{ex}}_{ab})$.}

\medskip

To apply this to CCC we'll take $g$ to be $\hat{g}$, the metric of the previous aeon, and $\tilde{g}$ to be $\check{g}$, the metric of the current aeon. Then we need $\phi$ to vanish at the crossover. Thus it is present in the 
previous aeon but fading to zero, while in the current aeon it starts very large (formally infinite).

\subsubsection{The Reciprocal Hypothesis applied to the case of FLRW}\label{SSS3.4.3}
The choices we made in section 3.3 have to be changed slightly to accord with the reciprocal hypothesis. We have
\[\hat{g}=d\hat{t}^2-\hat{a}^2d\sigma_k^2=\hat{\Omega}^2g,\;\;\check{g}=d\check{t}^2-\check{a}^2d\sigma_k^2=\check{\Omega}^2g,\]
so with $\hat{\Omega}=\hat{c}_1\hat{a},\check{\Omega}=\check{c}_1\check{a}$ we have
\[g=\frac{d\hat{t}^2}{\hat{c}_1^2\hat{a}^2}-\frac{1}{\hat{c}_1^2}d\sigma_k^2=\frac{d\check{t}^2}{\check{c}_1^2\check{a}^2}-\frac{1}{\check{c}_1^2}d\sigma_k^2,\]
whence $\hat{c}_1=\check{c}_1=c_1$ say  and 
\[d\tau=\frac{d\hat{t}}{c_1\hat{a}}=\frac{d\check{t}}{c_1\check{a}}.\]
The reciprocal hypothesis is
\[-1=\hat{\Omega}\check{\Omega}=c_1^2\hat{a}\check{a}\] 
Now by comparing with section 3.3, we'll have successive aeons diffeomeophic if $c_1^2=c=\sqrt{3\mu/\lambda}$. This calculation has also given us the metric of $\scri^+$ as $\frac{1}{c_1^2}d\sigma_k^2$ so the scalar 
curvature of $\scri^+$ is $6kc_1^2$ and the scalar curvature of the cross-over metric $g$ is minus this.

\subsection{The VBE and fading rest mass}\label{SS3.5}
In \cite{p1} Penrose makes a speculation which I don't think is essential to the CCC picture but is interesting and provocative in its own right. He contemplates the far future of a universe with positive $\lambda$. In a 
classic article \cite{fd} the remote future was considered by Freeman Dyson, before positive $\lambda$ became the consensus, and much of what he said must still hold up. His time-line for the far future 
had stars disappearing by $10^{14}$ years, 
galaxies by $10^{19}$ years and black holes by $10^{100}$ years (by Hawking evaporation -- recall that a black hole with mass $NM_\odot$ evaporates in a time of about $10^{67}N^3$ years). 
After that, it depends if there is a lower limit to the mass of a black hole: if the Planck mass is a lower limit then ordinary 
matter will spontaneously 
collapse to black holes on a time-scale of $10^{10^{26}}$ years, and these will then evaporate, but if there is no lower limit that process is more rapid and Dyson refers to \cite{z1} where a time-scale of 
$10^{26}$ years for all matter to collapse to black holes is suggested, and these then evaporate. In this last case of Dyson's scenarios for the remote future, after these black holes have gone, 
one has a universe containing only massless particles and gravitation. 

This is also the future that Penrose argues for but by a different physical mechanism, a kind of reverse Higgs mechanism: the Higgs mechanism is 
supposedly the process which gives rest mass to elementary particles at some stage in the early universe, and Penrose's speculation is that it turns off rest mass again at some late time. Evidently this needs `new physics', 
so it is striking that Dyson had a similar picture by a different route.

Once all particles are massless (or are 
once again massless) then there can be no clocks, and proper time disappears from the world. This is partly Penrose's response to what he called \emph{the VBE}: if the galaxies are gone by $10^{19}$ years or so then 
nothing much happens in the 
universe apart from occasional black hole mergers until the the supermassive black holes from galactic nuclei evaporate and this will take $10^{100}$ years -- this waiting characterises the Very Boring Era (\cite{p1}). If there are 
no clocks and therefore no 
proper time then it won't seem so long. This can also be seen as the resolution of an apparent paradox: the proper time until $\scri^+$ is infinite but how in the physical world 
can you have a completed infinity before the next aeon? The resolution 
is that proper time loses its physical significance -- there is a finite amount of conformal time before $\scri^+$.

\subsection{Physical consequences}\label{SS3.6}
If CCC is to be part of physics then it needs to make predictions which can be confirmed. While CCC is agnostic about inflation, the default position would be to seek to manage without it -- Penrose has argued against the 
inflation consensus for many years (see for example the discussion in \cite{p4}) and it is easier these days to find mainstream articles critical of inflation (e.g. \cite{isl}). 
A good argument that one can manage without it would be to derive the density perturbation in the early universe within CCC and without calling on inflation. On the face of it this could be done since 
there is an epoch of exponential expansion in CCC but it arises at the end of the previous aeon rather than very early in the current one -- can one quantitatively support the assertion that 
`inflation happened before the Bang'? -- but it hasn't so far been done. The emphasis has been on looking for other effects which come through from the previous aeon. The crossover surface is not a barrier to massless particles 
so that photons and gravitons should come through, and one might be able to detect the presence of a previous aeon by effects in electromagnetic or gravitational radiation. 

\medskip

We'll briefly mention primordial magnetic fields. There are magnetic fields at all scales in the universe \cite{lw}. Magnetic fields within galaxies are explicable given primordial magnetic fields to act as seeds for dynamo 
amplification processes \cite{kul}. Between galaxies there are known lower limits on  magnetic fields \cite{han} and these intergalactic fields may be primordial. The occurence of primordial magnetic fields may be 
attributed to phase changes in the early universe or to inflation see e.g. \cite{ksub}, \cite{lw}, but they could instead come through from the previous aeon. 

As a toy example, in the FRW metric
\[\hat{g}=dt^2-a^2(dx^2+dy^2+dz^2)\]
consider the 2-form
\[{\bf{B}}=Bdx\wedge dy\mbox{  so that } *{\bf{B}}=\frac{B}{a}dt\wedge dz.\]
This 2-form is evidently closed and co-closed so it solves Maxwell equations and corresponds to a Maxwell tensor $F_{ab}$ with $F_{ab}u^b=0$ where $u^a$ (as usual) is the fluid velocity or Hubble flow. Thus it's a pure 
magnetic field, and it's clearly smooth as a 2-form through the Bang, though for example, the norm $F_{ab}F^{ab}=2B^2a^{-4}$ isn't. Also one can calculate the energy-momentum tensor and obtain
\[T_{ab}u^b=\frac{B^2}{8\pi a^4}u_a.\]
One might therefore define $\frac{B^2}{8\pi a^4}$ as the energy density, which would go to zero faster than the volume diverges towards $\scri^+$. However if one uses 
instead the conformal Killing vector $X^a=au^a$ then there is a \emph{conserved} current
\[\tilde{J}_a=T_{ab}X^b=\frac{B^2}{8\pi a^3}u_a,\]
(foreshadowed in section \ref{SSS1.3.1}) and the integral over a comoving 3-volume is constant through the cross-over. 
Penrose has suggested looking for B-modes in the CMB in regions on the sky where the circles that we come to in the next section are densest (?ref) since magnetic fields in the previous aeon would be strongest 
inside superclusters and these regions of dense circles may be interpreted as superclusters hitting $\scri^+$.


\subsubsection{Circles in the sky}\label{SSS3.6.1}
Late in the previous aeon, all stars and galaxies should have gone but, before they've evaporated, there should be a population of supermassive black holes, relics of galactic clusters and superclusters. There will be 
mergers between these, releasing large bursts of gravitational radiation -- the black holes themselves could have masses of $10^{10}$ $M_{\odot}$  and as much as 40\% of the mass could be radiated -- and this burst will be of 
sufficiently short duration that it could be treated as a $\delta$-function wave supported on the light cone of the emission event. This wave will pass through the cross-over surface but immediately interact with the hot 
early universe in the next aeon. This will diffuse out the energy into a region between two concentric spheres and produce an inhomogeneity on the last-scattering surface of the CMB 
which will appear as an annulus of inhomogeneity on the 
intersection of the past light cone of an observer now with the last-scattering surface, which is essentially the CMB sky. This circular feature might be detectable by statistical properties: it might have detectably different 
mean temperature from the background; it might have detectably different temperature variance from the background.   

In a series of papers, Penrose and Gurzadyan have described a search for circles of lower variance, \cite{gp1,gp2,gp3,gp4}. The method is straightforward: choose a (large) set of centres, and for each centre a set of radii 
and widths for the annulus; 
then plot the temperature variance of the annuli, calculated from published CMB surveys, with a threshold for significance. To assign a measure of statistical significance repeat the process with a model sky 
constructed artificially 
but with the same statistical properties as the actual sky. Penrose and Gurzadyan have found large numbers of statistically significant circles, including sets of concentric circles where the same point on the sky is the 
centre of as many as four. 
However the statistical significance has been denied by several other groups: \cite{dcs,ew,haj,we}. The dispute centres on the way the artificial, comparison 
skies are constructed. The sets of concentric circles were unexpected but can be argued 
to correspond in the model to a supermassive black hole at rest with respect to the Hubble flow and into which a succession of smaller black holes falls. 

There is an interesting point emphasised in \cite{gp4} about the 
centres of concentric 
sets: the circles are identified by having low variance in temperature, but once identified their mean temperature can be calculated and the centre indicated by a blue or red dot on the sky, according as this mean is lower or 
higher than the ambient (see figure 2 in \cite{gp4}). The distribution of red and blue dots on the sky is strikingly inhomogeneous. To understand the dots it is useful to think in 3 dimensions: our past light cone meets 
the last-scattering surface in a 2-sphere and the future light cone of an imagined event in the previous aeon also meets the last scattering surface in a 2-sphere, which one can think of as expanding; 
these two 2-spheres meet in a circle; if the centre of the expanding 2-sphere is inside our 2-sphere then the shock wave is moving away from us, the temperature will be red-shifted and therefore lower and this 
event will get a blue dot; if the centre is outside then the shock wave is moving towards us, the temperature will be blue-shifted and therefore higher and this event 
will get a red dot. In the model then a clump of red dots represents 
a centre of activity outside the region bounded by our 2-sphere and a clump of blue dots represents a centre of activity inside our 2-sphere. If the circles, and therefore the distribution of their centres, are 
statistical artefacts then it is 
hard to see why such a coherent picture emerges -- why are there clumps of blue dots and clumps of red dots at all, rather than a mixed jumble of red and blue dots?

\medskip

In a different series of papers, Meissner, Nurowski and collaborators \cite{mnr,amn,amn2} have described a search for circles with anomalous \emph{mean} temperature. The method of search is similar 
but the statistical analysis is quite different and uses a nonparametric test due to Meissner \cite{m1}. In \cite{amn2} the method is modified: the statistic calculated is the difference between an average temperature 
over an inner ring and an average temperature over an outer ring, the two rings contiguous and forming a single wider ring. This follows a suggestion by Penrose that there should be a temperature 
profile of a sharp rise and slow decline across these rings (so this test statistic should be \emph{negative}. 
Both this series of papers and the Gurzadyan-Penrose series also tested the effect of `twisting' the observed sky. This, in terms of spherical polar coordinstes, 
is the (conformal) transformation
\[(\theta,\phi)\rightarrow(\tilde\theta,\tilde\phi)=(\theta,\phi+S\theta)\]
of the sky, for varying choices of the real parameter $S$ and would be expected to disrupt genuine rings but not artefactual ones. Both sets of authors find that the number of detected rings indeed declines sharply with 
increasing $S$.

One difference between the two sets of work is that the rings (strictly speaking, \emph{annuli}) of the Polish group are significantly wider than those of Gurzadyan-Penrose. 
 
\subsubsection{Shock waves through the cross-over surface}\label{SSS3.6.2}
We'll briefly describe a model for the generation of these circles by delta-function Ricci-curvature shock-waves from a source in the previous aeon, with the FLRW metric. Choose the source to be at the origin 
and put $k=0$ for convenience. We want to regard the spherically-symmetric shock-wave as a perturbation of the FLRW metric so that section \ref{SSS3.4.2} fixes 
conventions. The metrics in the two aeons are
\[\hat{g}=d\hat{t}^2-\hat{a}^2d\sigma^2,\;\;\check{g}=d\check{t}^2-\check{a}^2d\sigma^2,\]
with
\[d\sigma^2=dr^2+r^2(d\theta^2+\sin^2\theta d\phi^2).\]
The fluid content will be assumed to be radiation fluid with (unit, time-like) velocity vectors in the two aeons as
\[\hat{u}=\partial_{\hat{t}},\;\;\check{u}=\partial_{\check{t}}.\]
For a radiation fluid the conservation equation is solved for the densities by
\[\kappa \hat{\rho}_\gamma=\hat{\mu}/\hat{a}^4,\kappa\check{\rho}_\gamma=\check{\mu}/\check{a}^4\]
as before, with $\kappa=8\pi G/3$.

The shock wave is supported on the light cone of the origin at a certain time. This is best done in terms of the null coordinate
\[u=\tau-r\]
where $\tau$ is conformal time. Suppose the energy momentum tensor of the shell in the previous aeon is
\[\hat{T}^{(S)}_{ab}=\delta(u-u_E)\hat{F}(\hat{t},r)\hat{\ell}_a\hat{\ell}_b,\]
where the superscript $S$ is to distinguish this from the background $\hat{T}_{ab}$ which is the radiation fluid, $u_E=\tau_E$ which is the conformal time at emission, and
\[\hat{\ell}^a\partial_a=\partial_{\hat{t}}+\frac{1}{\hat{a}}\partial_r=\frac{1}{\hat{a}}(\partial_{\tau}+\partial_r),\]
which is the null generator of the light cone of the origin, normalised against the fluid velocity.

The conservation equation implies
\[r^2\hat{a}^4\hat{F}=\frac{\hat{m}}{4\pi}=\mbox{  constant}\]
so that $\hat{F}$ can be thought of as proportional to energy or mass per unit area on the spherically-symmetric expanding shock. 
There are two ways to interpret the constant $\hat{m}$. For one way, consider the conformal Killing vector $X^a=\hat{a}\hat{u}^a$ (it's easy to check that this \emph{is} a conformal Killing vector) and define the current
\[\hat{J}_a=\hat{T}^{(S)}_{ab}X^b=\delta(u-u_E) \hat{F}\hat{a}\hat{\ell}_a.\]
This is automatically conserved, and when integrated over a surface $\Sigma$ of constant $\hat{t}$ gives
\[\int_\Sigma \hat{J}_a\hat{u}^ad\Sigma=\int_\Sigma\delta(\tau-r-u_E)\hat{F}\hat{a}^4 r^2\sin\theta drd\theta d\phi=\hat{m},\]
so that $\hat{m}$ is the conserved quantity defined from this conserved current. This may not be the right definition of \emph{total energy in the shell} and it may be one should use $\hat{u}^a$ in place of $X^a$, 
in which case the energy in the shell is
\[\hat{E}:=\int_\Sigma \hat{T}^{(S)}_{ab}\hat{u}^a\hat{u}^bd\Sigma =4\pi r^2\hat{a}^3\hat{F}=\frac{\hat{m}}{\hat{a}}\]
which is not conserved: it's $\hat{E}=\hat{m}/\hat{a}_E$ initially but tails away to zero as the previous aeon expands, then jumps to infinity as the Bang is crossed and then decreases again.

The shell passes through to the next aeon and there will be corresponding checked quantities. How do we match them? The correct rescaling to preserve the conservation equation is
\[\mbox{  if  }\hg_{ab}=\Theta^4\cg_{ab}\mbox{   then   }\hT_{ab}=\Theta^{-4}\cT_{ab},\]
and we'll adopt this, which leads to $\hF=\Theta^{-8}\cF$ and then $\hm=\cm$. Now the choice in the definition of energy in the shell noted above is crucial: 
if the energy in the shell in the present aeon is $\cm=\hm$ then it's constant through the crossover; if rather 
it is $\check{E}:=\cm/\ca=\hm/\ca$ then it jumps up to infinity as it goes through, and then decays as the scale factor grows.

We want to calculate the density perturbation at the last-scattering surface in the current aeon due to a shock like this. When the shock comes through $\scri^+$ it will interact with the ambient hot radiation fluid 
and spread out. We look at this next.

\subsubsection{Width of the annuli and energetics}\label{SSS3.6.3}
The last-scattering surface is located at redshift $z=1089$ (\cite{bau}), or equivalently $S=1/1090$ with $S=a/a_0$ as in section \ref{SSS2.1.1}. We use this in (\ref{tc3}) to obtain $a_0H\tau_{LS}=0.052$ and in 
(\ref{tc4}) to obtain $t_{LS}= 3.68\times 10^5$ years, which is the accepted value (\cite{bau}) (note here we've taken the origin in $\tau$ and $t$ to be at the cross-over). 
The null cone of the spatial origin at $t=0=\tau$ is $\tau=r$, which occupies a sphere of coordinate radius $\tau_{LS}$ on 
the last-scattering surface. Our past light-cone meets the last-scattering surface in a sphere of coordinate radius $\tau_0-\tau_{LS}$ 
which will subtend an angle $2\delta\theta$ at us, where $\tan\delta\theta=\tau_{LS}/(\tau_0-\tau_{LS})=0.052/(2.66-0.052)=0.020$. This is small enough that we may approximate 
$\delta\theta=\tan\delta\theta=0.020$: a point on the cross-over 
surface can influence a region of this angular radius on the intersection of our past light-cone with the last-scattering surface. This is essentially the \emph{horizon problem}: $.04$ radians is a little over 2 degrees 
so that points on the CMB sky further apart than this have not been in causal contact, in the sense that their causal pasts do not overlap, \emph{since the Bang}. However, in CCC, their causal pasts do overlap but in the 
previous aeon. This is how CCC solves the horizon problem.

\medskip 

The shock-wave envisaged in the last section meets the crossover in a sphere of radius $-\tau_E$ ($\tau_E$ is negative as it refers to an event in the previous aeon). This sphere spreads out in the next aeon but by 
causality it is confined to the region between spheres 
of radii $-\tau_E\pm\tau_{LS}$. Exactly how wide this ring is will depend on the details of the diffusion of energy in the hot early universe. Its actual width shouldn't be more than $.04$ radians but the geometry 
of the intersection of this region with our past light cone can make it appear wider.

We obtain the size of the annulus on the last-scattering surface by using (\ref{tc7}) and values from section \ref{SSS2.1.1}:
\[\tan\theta=\frac{\tau_{LS}-\tau_E}{\tau_0-\tau_{LS}}=\frac{a_0H\tau_{LS}-a_0H\tau_E}{a_0H(\tau_0-\tau_{LS})}=\frac{0.052+1.31e^{-Ht_E}}{2.61},\]
where $t_E$ is proper-time of emission in the previous aeon, which we are assuming is essentially the same as the current one, and the factor $1.31$ is the approximate value of $\alpha^{-1/3}$.

Recall that at the present time $Ht_0\sim 0.82$, and the term in $t_E$ can be neglected when $Ht_E\geq 4.5$ so decent size circles, in the range $1.5^o<\theta<12^o$ will only be formed with $1<Ht_E<4.5$. Later than 
that the circles may be more like discs.

To get an order of magnitude we consider the smallest possible circle. This has radius $\tau_{LS}$ (since the light cone of the origin at $\tau=0$ has equation $r=\tau$) 
and the mass inside a sphere of this radius due to the background density is 
$M=\frac{4\pi}{3}a_{LS}^3(\tau_{LS})^3\rho$. The cosmological density, following section \ref{SSS2.1.1} is
\[\rho=\frac{A}{a_{LS}^3}+\frac{B}{a_{LS}^4}=(\alpha+\beta\frac{a_0}{a_{LS}})\left(\frac{a_0}{a_{LS}}\right)^3\frac{H^2}{\kappa}.\]
Also $a_0H\tau_{LS}=0.052$ as calculated above and 
\[(\alpha+\beta\frac{a_0}{a_{LS}})=0.45+1.4\times 10^{-4}\times 1090=0.60\]
so that
\[M=\frac{4\pi}{3}(0.052)^3\times 0.60\frac{1}{H\kappa}=7\times 10^{18}M_\odot.\]

The most violent events imaginable late in the previous aeon (or indeed anywhere) are mergers of supermassive black holes. There are currently believed to be black holes of mass $10^{10}M_\odot$ in the universe 
but these do seem to be the largest and there 
are arguments in the literature that these may be the largest possible: see \cite{nt}, \cite{ssh}. If two black holes collide then as much as $40\%$ of the rest mass can be emitted as gravitational radiation so the most extreme events 
should emit about $10^{10}M_\odot$. If the shock wave considered here dumps about $10^{10}M_\odot$ of energy into the sphere under consideration 
then that's a fraction $7\times 10^{-8}$ of the background, while the scale of actual density perturbations at last-scattering is usually said to be 
$\delta\rho/\rho\sim 10^{-5}$, so this is rather low. The fraction increases if the energy of 
the shock is concentrated closer to the surface of the sphere, rather than distributed all across it. A detailed calculation of the scattering process in the earlier universe would be needed to get this right. 
The other view on what energy should be is essentially ruled out as that would give a mass $10^{10}a_E/a_{LS}M_\odot$ dumped in the sphere, and the ratio $a_E/a_{LS}$ 
can be vast.

\medskip


\subsection{To do}\label{SS3.7}
The outstanding problems with CCC seem to me to require more physical cosmology rather than more mathematical cosmology. One wants a detailed model of the physical processes around the cross-over surface to answer questions like:
\begin{itemize}
 \item can `circles in the sky' be made to work? Are the events envisaged (namely super-massive black hole mergers in the previous aeon) of the right scale of energy and the right frequency of occurence to produce the circles?
 \item can magnetic fields really come through or are thet damped out on the `hot' side?
 \item is it possible to obtain the observed spectrum of density perturbations from CCC? (This is widely regarded as the remaining great achievement of theories of inflation.)
 \item can Penrose's suggestion of dark matter as `erebons' \cite{p7} be justified?
\end{itemize}

\end{document}